\documentclass[nofootinbib,twocolumn,preprintnumbers]{revtex4-1}
\usepackage{amsmath,amsthm,amssymb,multirow,psfrag}
\usepackage{epsfig}
\usepackage{color}

\graphicspath{{./Figures/}}

\begin{document}

\def\lsim{\mathrel{\rlap{\lower4pt\hbox{\hskip1pt$\sim$}}
    \raise1pt\hbox{$<$}}}
\def\gsim{\mathrel{\rlap{\lower4pt\hbox{\hskip1pt$\sim$}}
    \raise1pt\hbox{$>$}}}
\newcommand{\vev}[1]{ \left\langle {#1} \right\rangle }
\newcommand{\bra}[1]{ \langle {#1} | }
\newcommand{\ket}[1]{ | {#1} \rangle }
\newcommand{\ev}{ {\rm eV} }
\newcommand{\kev}{{\rm keV}}
\newcommand{\mev}{{\rm MeV}}
\newcommand{\gev}{{\rm GeV}}
\newcommand{\tev}{{\rm TeV}}
\newcommand{\mpl}{$M_{Pl}$}
\newcommand{\mw}{$M_{W}$}
\newcommand{\Ft}{F_{T}}
\newcommand{\Zparity}{\mathbb{Z}_2}
\newcommand{\BLambda}{\boldsymbol{\lambda}}
\newcommand{\be}{\begin{eqnarray}}
\newcommand{\ee}{\end{eqnarray}}
\newcommand{\met}{\;\not\!\!\!{E}_T}

\title{Dark Matter and Vector-like Leptons From Gauged Lepton Number}
\author{{\bf Pedro Schwaller$\, ^{a,b}$, Tim M.P. Tait$\, ^{c}$, Roberto Vega-Morales\,$^{d,e}$}}
  
\affiliation{
$^a$ HEP Division, Argonne National Laboratory, Argonne, IL, USA\\
$^b$ Physics Department, University of Illinois-Chicago, Chicago, IL, USA\\
$^c$ Physics Department, University of California-Irvine, Irvine, CA, USA\\
$^d$ Fermi National Accelerator Laboratory (FNAL), Batavia, IL, USA,\\
$^e$ Department of Physics and Astronomy, Northwestern University, Evanston, IL, USA}

\begin{abstract}
We investigate a simple model where Lepton number is promoted to a local $U(1)_L$ gauge symmetry which is then spontaneously broken, leading to a viable thermal DM candidate and vector-like leptons as a byproduct. The dark matter arises as part of the exotic lepton sector required by the need to satisfy anomaly cancellation and is a Dirac electroweak (mostly) singlet neutrino. It is stabilized by an accidental global symmetry of the renormalizable Lagrangian which is preserved even after the gauged lepton number is spontaneously broken and can annihilate efficiently to give the correct thermal relic abundance. We examine the ability of this model to give a viable DM candidate and discuss both direct and indirect detection implications. We also examine some of the LHC phenomenology of the associated exotic lepton sector and in particular its effects on Higgs decays.
\end{abstract}
\preprint{FERMILAB-PUB-13-127-T, NUHEP-TH/13-2}
\preprint{UCI-HEP-TR-2013-07, ANL-HEP-PR-13-23}


\maketitle


\section{Introduction}
\label{sec:Intro}
With the recent discovery of a new resonance with standard model (SM) Higgs like 
properties~\cite{:2012gk,:2012gu} the final piece of the SM appears to be in place. It is well known, 
however, that there are questions for which the SM has no answer and beyond the standard (BSM) 
physics is needed. Chief among these questions is the nature of dark matter (DM) and the 
mechanism which makes it stable. It is also well known that the renormalizable SM Lagrangian 
possesses an (anomalous)
accidental global symmetry associated with the conservation of overall lepton 
number. If one allows for higher dimensional operators, lepton violating interactions can occur
at dimension five, but to date no such processes 
(with the possible ambiguous exception of neutrino masses)
have been observed experimentally~\cite{Beringer:1900zz}. This is perhaps an indication 
that lepton number is a more fundamental symmetry which prevents the generation of SM lepton 
number violating operators. In this work, 
we connect the apparent lack of lepton number violation 
to the stability of thermal relic dark matter, by deriving both from a $U(1)_L$ 
gauge symmetry associated with lepton number. 

Gauging lepton number is attractive for both phenomenological as well as theoretical reasons and the 
possibility of lepton number (and also baryon number) as a local gauge symmetry was first explored in~\cite{Rajpoot:1987yg,Foot:1989ts}. However, the first complete and consistent model of gauged lepton number (and baryon number) was not explored until more recently in~\cite{FileviezPerez:2010gw} with numerous variations following~\cite{Dulaney:2010dj,FileviezPerez:2011pt,FileviezPerez:2011dg,Duerr:2013dza,Chao:2010mp,Dong:2010fw}. Here we explore a specific realization where the DM arises as part of the exotic lepton sector required by gauging lepton number and the attendant need to cancel anomalies. We study in particular the DM and LHC phenomenology of this construction.

The DM candidate is a Dirac electroweak (mostly) singlet neutrino stabilized by an accidental 
global symmetry of the renormalizable Lagrangian which is preserved even after
lepton number is spontaneously broken. As we will see, as a byproduct of the lepton breaking 
mechanism and the requirement of a viable DM candidate, one also obtains a set of vector-like 
leptons which can have interesting phenomenology at the LHC through either direct production or 
through modifications of Higgs decays to SM particles. 

We extend the SM gauge group to $SU(3)_c\otimes SU(2)_W\otimes U(1)_Y\otimes U(1)_L$ 
where the SM leptons are charged under $U(1)_L$. The anomalous $U(1)_L$ requires us to add a 
new set of leptons with the appropriate quantum numbers to cancel anomalies. 
Typically, $U(1)_L$ is spontaneously broken by the vacuum expectation value of a SM singlet 
scalar in such a way that Majorana masses can be generated for the right-handed neutrinos, 
(whose presence is required by anomaly cancellation~\cite{FileviezPerez:2010gw}).
Such constructions allow for a simple realization of the well known `see-saw' mechanism of neutrino mass generation,
but do not contain viable dark matter candidates without additional assumptions or particle content.

Here, motived by the desire for a thermal DM candidate, we choose to break lepton number 
with a SM singlet scalar carrying $L=3$. This leads to a remnant global $U(1)$ symmetry preventing decay of the lightest 
new lepton which stabilizes the DM candidate. This global symmetry is a 
consequence of the gauge symmetry and particle content of the model and does not need to be 
additionally imposed. It also ensures that the model is safe from dangerous flavor violating 
processes which are highly constrained by experiment. An automatic consequence of this construction is that one also obtains a new 
generation of vector-like (with respect to the SM) leptons after the spontaneous breaking of lepton 
number. This type of lepton spectrum has garnered recent interest in the context of modifications to the Higgs decay into 
diphotons~\cite{Joglekar:2012vc,ArkaniHamed:2012kq,Batell:2012zw,Arina:2012aj,Moreau:2012da,Feng:2013mea,Joglekar:2013zya}
and was also recently shown to be useful for baryogenesis~\cite{Davoudiasl:2012tu,Perez:2013nra}.

The organization of this papers is as follows.  In Sec.~\ref{sec:gaugedL} we briefly review the gauging of lepton number and 
cancellation of anomalies. We also discuss the details of the lepton breaking mechanism as well as the particle content and 
Lagrangian. In Sec.~\ref{sec:darkmatter} we discuss the DM candidate and stability and obtain the relic abundance for a 
range of DM masses. We also examine the direct and indirect detection prospects. In Sec.~\ref{sec:lhc_pheno} we discuss 
constraints as well as LHC phenomenology and examine the effect of the vector-like leptons on the Higgs to 
diphoton rate. We present our conclusions and an overview of possible future work in Sec.~\ref{sec:conclusions}.

\section{The Model}
\label{sec:gaugedL}

The SM gauge group is extended to 
$SU(3)_c \otimes SU(2)_W\otimes U(1)_Y\otimes U(1)_L$ 
where $L$ represents the lepton charge. We restrict ourselves to the minimal particle content 
consisting of a set of anomaly-canceling exotic leptons, plus the new gauge field and a 
SM singlet scalar which breaks lepton number spontaneously.
In principle, this theory is UV-complete up to large energies, and we restrict ourselves to
considering renormalizable interactions. 
We discuss each of these  ingredients, including the interactions, below.

\subsection{Anomaly Cancellation}
\label{sec:Anomaly}

The anomalies introduced when gauging lepton number and various ways to cancel them 
with the addition of new fermions are discussed in detail 
in~\cite{FileviezPerez:2010gw,Dulaney:2010dj,FileviezPerez:2011pt}.
All options include three generations of right-handed singlet neutrinos 
($\nu_{Ri}$, considered as part of the SM) with quantum numbers 
$\nu_{Ri}\equiv (1,0,1)$ under $(SU(2)_W,  U(1)_Y, U(1)_L)$ and 
$i=e,\mu,\tau$.  We define all SM leptons to have $L=1$. In addition to $\nu_{Ri}$, one must add 
new electroweak doublet and singlet leptons to cancel the gauge anomalies. 
There are several options; here we focus on a simple construction making use of two exotic 
generations of chiral fermions which together form a vector-like set under the SM gauge 
group~\cite{FileviezPerez:2011pt}, ensuring that anomaly cancellation in the SM gauge factors
is preserved. 
The first set of new fermions is a sequential fourth generation of leptons
carrying lepton number $L=L^\prime$,
\begin{eqnarray}
\label{eqn:leps1}
&&\ell_{L}^\prime \equiv (\nu_{L}^\prime~e_{L}^\prime) \equiv (2,-1/2,L^\prime),\nonumber \\
&& e_{R}^\prime \equiv (1,-1,L^\prime),~\nu_{R}^\prime \equiv (1,0,L^\prime) .
\end{eqnarray}
The second is a mirror set of opposite chirality with lepton number 
$L=L^{\prime\prime} = L^\prime + 3$,
\begin{eqnarray}
\label{eqn:leps2}
&&\ell_{R}^{\prime\prime} \equiv (\nu_{R}^{\prime\prime}~e_{R}^{\prime\prime}) 
\equiv (2,-1/2,L^{\prime\prime}), \nonumber\\
&&e_{L}^{\prime\prime} \equiv (1,-1,L^{\prime\prime}),~\nu_{L}^{\prime\prime} 
\equiv (1,0,L^{\prime\prime}) , 
\end{eqnarray}
where the condition,
\begin{eqnarray}
\label{eqn:anom_condition}
L^\prime - L^{\prime\prime} = -3 
\end{eqnarray}
is required by anomaly cancellation. The addition of two sets of chiral fermions carrying lepton 
number which together form a vector-like set under the SM also avoids 
the need to add new quarks to cancel anomalies, although scenarios with exotic quarks are also 
interesting and have been explored in the context of gauged baryon 
number~\cite{Dulaney:2010dj,FileviezPerez:2010gw,FileviezPerez:2011pt}. The particle content in Eqs.(\ref{eqn:leps1}) and (\ref{eqn:leps2}) is similar to that obtained in~\cite{Duerr:2013dza} where baryon number is also gauged and one obtains a vector-like set of `lepto-quarks' as well as a potential DM candidate. Here we focus on the case where only lepton number is gauged and examine the phenomenology in detail. 

\subsection{Gauge and Higgs Sector}
\label{sec:gauge_higgs_sector}

The gauging of lepton number will introduce a new spin-1 vector boson which we label $Z_L$. In 
addition to the usual Abelian vector field kinetic terms, the $U(1)_L$ gauge field will have
interactions,
\begin{eqnarray}
\label{eqn:ZLag}
&&\mathcal{L} \supset (D^\mu\Phi)^\dagger(D_\mu\Phi) + \frac{\epsilon}{2} Z^{\mu\nu}_L B_{\mu\nu} \nonumber \\
&& +~\bar{l}^\prime_L D_\mu \gamma^\mu l^\prime_L +\bar{l}^{\prime\prime}_R D_\mu \gamma^\mu l^{\prime\prime}_R 
+\bar{l}_i D_\mu \gamma^\mu l_i \,,
\end{eqnarray}
where $D^\mu = \partial^\mu + ig^\prime L Z_L^\mu$ with $L$ the lepton number assignment for 
a particular field.  $\Phi \equiv (1,0,L_\Phi)$ is the SM 
singlet scalar carrying lepton number whose 
\emph{vev} ($v_\phi$) breaks the $U(1)_L$ spontaneously.
The index $i = e, \mu, \tau$ runs over all SM leptons while $l = \ell, e, \nu$ where $\ell$ is an $SU(2)$ doublet and $e,\nu$ are 
singlets. Note  there is no $\delta M^2 Z_{L\mu}Z^\mu$ term since $\Phi$ is not charged under the SM and the Higgs does not 
carry $L$.

The parameter $\epsilon$ encapsulates the degree of kinetic mixing between $U(1)_L$ and
$U(1)_Y$.  One can in principle impose $\epsilon = 0$ at tree level through symmetries, but in general it is a free parameter of the theory and is additively renormalized by loops of
leptons. While any value of $\epsilon$ at the weak scale can be engineered, the loop-induced piece is typically of order
$10^{-3}$, small enough to be consistent with experimental constraints without undue fine tuning.

After lepton and electroweak symmetry breaking $\epsilon$ also leads to $Z-Z_L$ mixing 
parameterized by~\cite{Babu:1997st},
\begin{equation}
\label{eqn:Zmixingangle}
\tan 2\xi = \dfrac{2 M_Z^2 s_{W} \epsilon\sqrt{1-\epsilon^2}  }
{M_{Z_L}^2 - M_Z^2 (1 - \epsilon^2) + M_Z^2 s_W^2 \epsilon^2} \ ,
\end{equation}
where $\xi$ is the $Z_L - Z$ mixing angle and $M_Z, M_{Z_L}$ are the masses. In the absence of mixing, 
$M_{Z_L} = L_\Phi g^\prime v_\phi$. As we will see, since this mixing is constrained to be small by direct searches for dark 
matter (with weaker constraints from precision measurements \cite{Dienes:1996zr, Babu:1997st, Hook:2010tw}) 
we take $M_Z, M_{Z_L}$ as the physical masses as well. 

In the Higgs sector the existence of $\Phi$ allows for an expanded scalar potential, 
\begin{eqnarray}
\label{eqn:scalarlag}
&&V(H, \Phi)  = -\mu_H^2 H^\dagger H + \lambda_H |H^\dagger H|^2\nonumber \\
&& - \mu_\Phi^2 \Phi^\dagger\Phi + \lambda_\Phi |\Phi^\dagger \Phi|^2 + \lambda_{hp} \Phi^\dagger \Phi H^\dagger H,
\end{eqnarray}
where $H \equiv (2,-1/2,0)$ is the SM Higgs doublet. 
Once lepton number is broken, the real component of $\Phi$ obtains a vacuum expectation 
value $\langle \Phi \rangle = v_\phi/\sqrt{2}$, while the Higgs boson $H$ obtains its 
own \emph{vev}, $\langle H \rangle = (0,v_h/\sqrt{2})$ to break the electroweak symmetry. 
The scale $v_\phi$ will be the only new dimensional scale introduced, with all of the 
other parameters being dimensionless couplings. 
We will see below in Sec.~\ref{sec:YukLag} that $L_\Phi = 3$ is preferred.

The presence of the `Higgs portal' coupling $\lambda_{hp}$ will generically lead to mixing between the real 
singlet components of $\Phi$ and $H$ parameterized by the mixing angle,
\begin{equation}
\label{eqn:mixingangle}
\tan 2\theta = \dfrac{\lambda_{hp} v_h v_{\phi}}{ \lambda_\Phi v_\phi^2 - \lambda_H v_h^2} \ .
\end{equation}
This mixing leads to the mass eigenstates,
\begin{equation}
\label{eqn:s1s2_definition}
\begin{array}{ccc}
h &=& c_\theta h_o  - s_\theta\phi_o \ , \\
\phi &=& s_\theta h_o + c_\theta\phi_o \ , \\
\end{array}
\end{equation}
where $\phi_o$ and $h_o$ are the gauge eigenstates and $\phi, h$ are the mass eigenstates with masses, 
\begin{eqnarray}
\label{eqn:Ms1}
&& m^2_{h,\phi} = 
\left(\lambda_H v_h^2 + \lambda_\Phi v_\phi^2 \right) \mp
\sqrt{ \left(\lambda_\Phi v_\phi^2 - \lambda_H v_h^2 \right)^2 + \lambda_{hp}^2 v_h^2
  v_\phi^2 }\,, \nonumber \\
\end{eqnarray}
where we have assumed  $m_\phi > m_h$ and defined $c_\theta = \cos\theta$, $s_\theta = \sin\theta$, etc. 
The coupling $\lambda_{hp}$ will also lead to a tree level shift in the Higgs quartic 
coupling~\cite{EliasMiro:2012ay}, which 
provides a mechanism for stabilizing the vacuum in the presence of the exotic charged 
leptons with large Yukawa couplings to the SM Higgs. 
It was shown to be a particularly efficient stabilization mechanism when 
$m_\phi \gg m_h$, even for small mixing angles~\cite{Batell:2012zw}.

\subsection{Global Symmetries and Breaking L}
\label{sec:global}

The two new sets of leptons along with the SM lepton sector comprise three separate sectors 
labeled by their lepton number $L = 1, L^\prime, L^{\prime\prime}$ for which global $U(1)$ 
symmetries can be associated. These global symmetries are each separately conserved by the SM 
and $U(1)_L$ interactions. 
Yukawa interactions (assuming $L_\Phi$ permits them) will break these symmetries in
realistic models, as discussed below.
A combination of precision electroweak,
collider, and direct detection constraints prohibit a stable lepton which carries electroweak charge.
Thus, couplings to the Higgs must not be too large and the DM can not receive its mass solely from 
the SM Higgs, leading to the need to generate an additional contribution to the DM mass which does not come from electroweak symmetry breaking. 

From these considerations one concludes that the SM singlets $\nu^{\prime}_R$ and 
$\nu^{\prime\prime}_L$ or some combination must compose the majority of the DM.   Majorana 
masses can be generated by choosing the lepton breaking scalar to carry $L_\Phi=2L^\prime$ 
or $L_\Phi=2L^{\prime\prime}$. However, this choice still leaves either $L^\prime$ 
or $L^{\prime\prime}$ unbroken meaning that the lightest lepton of the corresponding sector will be 
stable and only receive its mass from its couplings to the Higgs, which as discussed is ruled out by 
experiment. It is clear that in order to avoid a heavy stable lepton with unacceptably large couplings 
to the $Z$ or Higgs boson one must choose $L_\Phi$ such that it generates an interaction between 
the $L^\prime$ and $L^{\prime\prime}$ sectors. The anomaly cancellation condition of 
Eq.(\ref{eqn:anom_condition}) ensures that the only possibility is $L_\Phi=3$. 

\subsection{Yukawa Sector}
\label{sec:YukLag}

Given $L_\Phi = 3\,$, the Lagrangian for the Yukawa sector of the new leptons can be written,
\begin{eqnarray}
\label{eqn:yuklag}
&&\mathcal{L} \supset -c_{\ell} \Phi\bar{\ell}^{\prime\prime}_{R}\ell^{\prime}_{L} 
- c_e \Phi\bar{e}^{\prime\prime}_{L}e^{\prime}_{R} - 
c_{\nu} \Phi\bar{\nu}^{\prime\prime}_{L}\nu^{\prime}_{R}  - y^{\prime}_e H\bar{\ell}^{\prime}_{L}e^{\prime}_{R}
\nonumber \\
&& - y^{\prime\prime}_e H\bar{\ell}^{\prime\prime}_{R}e^{\prime\prime}_{L} 
- y^{\prime}_\nu \tilde{H}\bar{\ell}^{\prime}_{L}\nu^{\prime}_{R} - y^{\prime\prime}_\nu \tilde{H}\bar{\ell}^{\prime\prime}_{R}\nu^{\prime\prime}_{L} + \emph{h.c.} .
\end{eqnarray}
In general these couplings are complex, containing phases which can lead to CP violation, but for simplicity we assume all 
couplings in Eq.(\ref{eqn:yuklag}) are real (but see~\cite{Voloshin:2012tv,McKeen:2012av} for recent studies of CP violating 
effects on the diphoton rate coming from vector-like leptons). 
It is also clear from Eq.(\ref{eqn:yuklag}) that once $\Phi$ obtains a \emph{vev} the couplings $c_\ell$, $c_e$, and $c_\nu$ will 
lead to vector-like (with respect to the SM) masses for the exotic leptons. The new leptons will also receive mass contributions 
from electroweak symmetry breaking through the $y_{\nu,e}^{\prime}, y_{\nu,e}^{\prime\prime}$ couplings. Note also that unless 
$L^\prime, L^{\prime\prime} = 0\,,$ explicit Majorana masses for $\nu^\prime_R$ and $\nu^{\prime\prime}_L$  are not allowed 
nor will they be generated after lepton number breaking unless $L^\prime = - L^{\prime\prime} = -3/2$ (This case was considered explicitly in the context of gauged lepton and baryon number with vector-like `lepto-quarks'~\cite{Duerr:2013dza}). We avoid these choices 
in what follows.

In principle there may still be couplings between the exotic and SM leptons.  Since we have taken SM lepton number to 
be $L=1$, this implies that $L^\prime, L^{\prime\prime} \neq 1$ in order to avoid mixing with SM leptons which can lead 
to dangerous flavor changing neutral currents as well as the decay of the DM. If we choose $L^\prime = -4$, which 
fixes $L^{\prime\prime}=-1$ then, in addition to those in Eq.(\ref{eqn:yuklag}), one can also generate interactions between 
the SM and the new lepton sector given by,
\begin{eqnarray}
\label{eqn:lag1SM}
{\mathcal L} \supset y \Phi \bar{\nu}^{c\prime}_{R} \nu_{Ri} + h.c. .
\end{eqnarray}
Once $\Phi$ obtains a \emph{vev}, this will lead to mixing between the SM right-handed neutrinos, $\nu_{Ri}$ and the exotic 
right handed neutrino, $\nu^\prime_{R}$. This also implies that the exotic lepton sector can decay to the SM, thus 
eliminating this scenario as an explanation for dark matter. To summarize, in order to avoid mixing with the SM and ensure 
a stable DM candidate, we take $(L^\prime, L^{\prime\prime}) \neq (1,4), (-4, -1), (-2, 1)$. Furthermore, to avoid Majorana mass 
terms we also assume $(L^\prime, L^{\prime\prime}) \neq (0,3), (-\frac{3}{2}, \frac{3}{2}), (-3, 0)$. Thus our complete Yukawa 
sector Lagrangian is given by Eq.~(\ref{eqn:yuklag}) and $L^\prime$ can otherwise be any real number satisfying $L^\prime = -3 + L^{\prime\prime}$.

In the limit that the Yukawa couplings $c_i\rightarrow 0$, one recovers the global symmetries which separately 
preserve $L^\prime$, $L^{\prime\prime}$ and $L_{SM}$. 
As a result, $c_i \ll 1$ are technically natural, implying that vector-like masses for the new leptons
much smaller than $v_\phi$ are natural. 
We also note that small values of the $y_{\nu,e}^{\prime}, y_{\nu,e}^{\prime\prime}$, and $y_{\nu i}^{SM}$ Higgs Yukawa couplings are technically natural.

It is worth noting that Eq.~(\ref{eqn:yuklag}) is very similar to the Yukawa sectors proposed in a 
generic framework in~\cite{Batell:2012zw, Joglekar:2012vc}, but here
arises from $U(1)_L$ gauge invariance and anomaly cancellation. 
Only one new scale ($v_\phi$) is introduced, with the masses of the new fermions following from dimensionless couplings. 
Furthermore, the global symmetries needed to protect against dangerous mixing with SM leptons and insuring
the existence of a stable DM particle are guaranteed by $U(1)_L$ gauge invariance as opposed to 
being imposed by hand.  

\subsection{Experimental Constraints}
\label{sec:constraints}

Low energy experiments place a limit on the parameters which describe the $Z_L$ sector. Since the SM
Higgs does not carry lepton number and $\Phi$ is a SM singlet, there is no mass-mixing between $Z_L$ and the
SM electroweak interaction at tree level. Furthermore since $Z_L$ does not couple to quarks, direct search limits
from the LHC are rather weak, and the strongest limits are obtained from constraints on four-lepton
operators derived from LEP II data \cite{Carena:2004xs}; these require 
\begin{eqnarray}
v_\phi \geq 1.7~{\rm TeV},
\end{eqnarray} 
roughly independently of the value of $g^\prime$. 

\begin{figure}[t]
\includegraphics[width=0.45\textwidth]{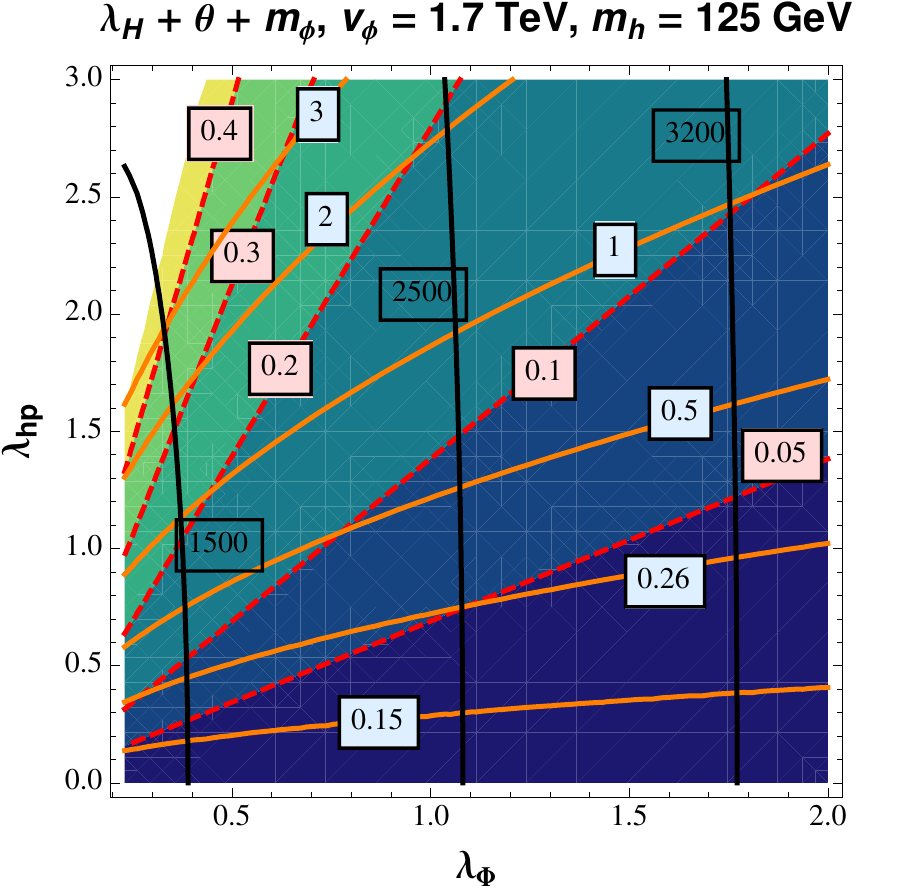}
\caption{Contours of Higgs mixing angle $\theta$(red-dotted), Higgs quartic coupling $\lambda_H$(orange-solid), and heavy scalar mass $m_\phi$ in GeV (black-solid) as defined in Eqs.(\ref{eqn:mixingangle}) Eq.(\ref{eqn:Ms1}) as a function of scalar couplings ($\lambda_{hp}, \lambda_\Phi$) in Eq.(\ref{eqn:scalarlag}).}
\label{fig:lam_vs_lam}
\end{figure}

This lower bound and the experimentally measured value of $m_h \simeq 125$~GeV constrains the quartic 
couplings in the scalar potential of Eq.(\ref{eqn:scalarlag}) through Eq.(\ref{eqn:mixingangle}) and (\ref{eqn:Ms1}). Fixing $v_\phi = 1.7$~TeV and $m_h = 125$~GeV we can then examine the scalar mixing angle $\theta$, the Higgs quartic $\lambda_H$, and the heavy scalar mass eigenstate $m_\phi$ as functions of the scalar couplings $\lambda_{hp}$ and $\lambda_\Phi$. In Fig.\ref{fig:lam_vs_lam} we show contours of $\lambda_H(\lambda_\Phi,\lambda_{hp})$ (solid-orange), $\theta(\lambda_\Phi,\lambda_{hp})$ (dotted-red), and $m_\phi(\lambda_\Phi,\lambda_{hp})$ (solid-black) 
in the $\lambda_{hp} - \lambda_\Phi$ plane.  As can be seen, values of $\theta \lesssim 0.1-0.2$ can be obtained for quartic couplings of $\mathcal{O}(1)$ and heavy scalar masses $\sim 2.5$~TeV. To obtain mixings as large as $\theta \sim 0.4$ requires $\lambda_H \sim 3$ and small $\lambda_\Phi \lesssim 0.5$ with $m_\phi \sim 1.5$~TeV. In general we find $m_\phi \gtrsim 1$~TeV for $v_\phi = 1.7$~TeV, possibly within reach of the LHC, but more likely too heavy to be produced directly.

\begin{figure}
\includegraphics[width=0.45\textwidth]{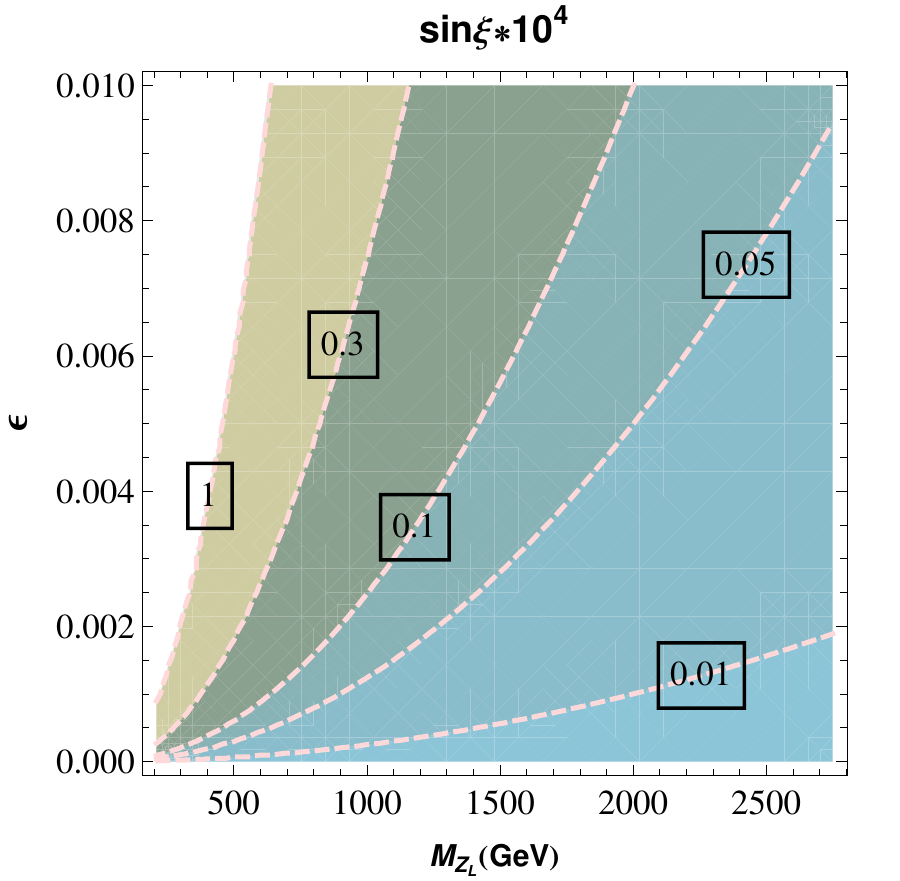}
\caption{Contours of the $Z-Z_L$ mixing angle $\sin\xi$ ($\times 10^{4}$) in the $\epsilon - M_{Z_L}$ plane (see Eq.(\ref{eqn:Zmixingangle})).}
\label{fig:ep_vs_mZL}
\end{figure}

Precision measurements on the $Z$-pole also constrain the degree of $Z_L$-$Z$ mass mixing.  Since this occurs at loop level
(through loops of the SM and exotic leptons as well as scalars),
it will typically be small enough ($\lesssim 10^{-3}$) for any $v_\phi$ consistent with the LEP II bound.  There are also
constraints (via $\sin\xi$ in Eq.(\ref{eqn:Zmixingangle})) on the kinetic mixing parameter
from direct detection~\cite{Aprile:2012nq}, 
which are comparable to the expected size induced by loops of leptons. Using Eq.(\ref{eqn:Zmixingangle}) we examine 
the $\epsilon - M_{Z_L}$ parameter space for typically allowed values of $\sin\xi \lesssim 10^{-4}$ over a 
range of $Z_L$ masses. In Fig.\ref{fig:ep_vs_mZL} we present contours of $\sin\xi \times 10^{4}$ in 
the $\epsilon-M_{Z_L}$ plane for small values of the kinetic mixing parameter $\epsilon$ as would be favored in 
theories where $\epsilon = 0$ at tree level as discussed in Sec.\ref{sec:gauge_higgs_sector}. We can see that for $M_{Z_L} \sim 1$~TeV one can obtain a $Z-Z_L$ mixing angle of $\sin\xi \sim 0.1\times 10^{-4}$ with a kinetic mixing of $\epsilon \sim 0.002$.

\subsection{Possible Extensions}
\label{subsec:extensions}
There are a number of possibilities for how one could extend this model or embed it into a more complete theory. For instance,with the need to break lepton number spontaneously, the question as to how one obtains $v_\phi$ naturally also arises. One 
could imagine embedding this model in a supersymmetric version as was done 
in~\cite{FileviezPerez:2011dg,FileviezPerez:2011pt,FileviezPerez:2012iw} for other gauged lepton number 
constructions. Another possibility is to have the scalar sector of this model arise as part of a set of goldstone bosons resulting from a strongly broken global symmetry~\cite{Gripaios:2009pe,Mrazek:2011iu}. 

Another possibility for generating natural values for not only $v_\phi$, but also the electroweak scale ($v_h$) is through 
dimensional transmutation where $v_\phi$ is generated radiatively~\cite{Coleman:1973jx}. This scale is then inherited by the SM 
through the `Higgs Portal' as done recently in~\cite{Englert:2013gz} for a hidden $U(1)$ gauge extension of the SM, but we 
leave it to a future study to explore this possibility. For the remainder of this study we simply set $v_\phi$ to its lower bound of $v_\phi = 1.7$ 
TeV.

One can also extend the theory to obtain $\epsilon = 0$ at tree level in Eq.(\ref{eqn:ZLag}) by positing that the $U(1)_L$ gauge symmetry arises out of a larger 
non-Abelian gauge symmetry which forbids $\epsilon \neq 0$ \cite{Carone:1995pu} and is broken at some high scale 
$\Lambda$ down to $U(1)_L$. Below the scale $\Lambda$, but above the lepton and electroweak breaking scales, 
loop corrections due to hyper-charged leptons vanish provided the leptons satisfy an orthogonality condition~\cite{Carone:1995pu},
\begin{equation}
\label{eqn:trace}
{\rm Tr} \left( L Y \right) = 0 .
\end{equation}
Combined with the anomaly cancellation constraint in Eq.(\ref{eqn:anom_condition}), 
 this would determine the exotic lepton numbers to be $ L' = -3$ and $L'' = 0$. 
Below $v_\phi$ and $v_h$ there will be loop induced (from both leptons and scalars) corrections which generate a kinetic mixing, but typically $\epsilon \ll 1$.

Note, that although we have only gauged lepton number, this is enough to prevent the dimension six operators of the form $\mathcal{L} \sim \frac{1}{\Lambda^2} qqq\ell$ (for appropriate lepton number assignment to the lepton breaking scalar) which might lead to proton decay. However, while baryon number violating operators at dimension six are forbidden, higher dimensional operators are still allowed since baryon number is not protected by a gauge symmetry. The leading operator that might mediate proton decay, 
\begin{eqnarray}
\mathcal{O} \sim \frac{c}{\Lambda^8} (qqq\ell) (\ell H)^2 \Phi^\dagger,
\end{eqnarray} 
first occurs at dimension twelve while $\Delta B = 2$ operators with $\Delta L = 0$ are allowed at dimension 9~\cite{Mohapatra:2009wp}, as in the SM. For $c\sim 1$ and scales $\Lambda \gtrsim \mathcal{O}(100)$~TeV the model considered here should be reasonably safe from the effects of these potentially dangerous operators. Of course one can extend this model to include gauged baryon number as well to prevent these operators~\cite{Duerr:2013dza}.

Finally is is worth mentioning that this model possesses many ingredients which may be helpful for explaining the baryon asymmetry of the universe. The current construction automatically contains new massive states as well as new interactions containing $CP$-violating phases. It would be interesting to explore whether or not it is capable of explaining this asymmetry as well as dark matter. Since the WIMP in this theory
is a Dirac fermion, there is potential to realize a theory with asymmetric dark matter. We leave it to future studies to explore these possibilities.

\section{Dark Matter}
\label{sec:darkmatter}
Here we examine the DM matter candidate in this model. We first discuss the stability which results from an accidental global symmetry of the Lagrangian and identify the DM as a heavy mostly singlet neutrino. This global symmetry is a consequence of the particle content and underlying lepton gauge symmetry, much in the same way that lepton number is an accidental global symmetry in the SM. We then discuss the various annihilation channels and calculate the relic abundance of the DM candidate to establish the allowed masses. We also discuss various other phenomenological features.

\subsection{DM Candidate and Stability}
\label{sec:DMcand}

We begin by examining the neutrino sector once $\Phi$ and $H$ obtain expectation values which gives,
\begin{eqnarray}
\label{eqn:DMyuklag}
&&\mathcal{L} \supset - \frac{c_\ell v_\phi }{\sqrt{2}} (1 + \frac{\phi_o}{v_\phi })  \bar{\nu}^{\prime\prime}_{R} \nu^{\prime}_{L} 
- \frac{c_\nu v_\phi }{\sqrt{2}} (1 + \frac{\phi_o}{v_\phi }) \bar{\nu}^{\prime\prime}_{L} \nu^{\prime}_{R}  \\
&& -\frac{y^{\prime\prime}_\nu v_h}{\sqrt{2}} (1 + \frac{h_o}{v_h})\bar{\nu}^{\prime\prime}_{R}\nu^{\prime\prime}_{L} 
- \frac{y^{\prime}_\nu v_h}{\sqrt{2}} (1 + \frac{h_o}{v_h}) \bar{\nu}^{\prime}_{L}\nu^{\prime}_{R} + \emph{h.c.} ,\nonumber 
\end{eqnarray}
leading to the mass matrix,
\begin{equation}
\label{eqn:nu_mass_matrix}
\mathcal{M}_\nu = \frac{1}{\sqrt{2}} \left( \begin{array}{cc}
c_\ell v_\phi & y^{\prime}_\nu v_h \\
y^{\prime\prime}_\nu v_h &   c_\nu v_\phi \\
\end{array} \right),
\end{equation}
which can be diagonalized using the singular value decomposition $\mathcal{M}_{\nu D} = U^\dagger_L \mathcal{M}_\nu U_R$, 
where $\mathcal{M}_{\nu D}$ is a diagonal mass matrix with positive mass eigenvalues $m_{\nu_X}$ and $m_{\nu_4}$. 

While the Yukawa couplings to $\Phi$ and $H$ break the global U(1) symmetries associated with $L'$ and $L''$ explicitly, there 
is a residual $Z_2$ symmetry under which all heavy leptons are odd and all SM leptons are even, which is preserved after 
spontaneous breaking of the lepton number and electroweak gauge symmetries. Assuming that the new charged leptons are 
heavier, this residual global symmetry guarantees the stability of the lighter of the two neutrino mass eigenstates,
opening up the possibility for dark matter.

In the limit where $y^{\prime}_\nu v_h, y^{\prime\prime}_\nu v_h \ll c_{\ell,\nu} v_\phi$, the mass eigenvalues are 
approximately given by
\begin{eqnarray}
\label{eqn:DM_masses}
&&m_{\nu_X} \approx \frac{1}{\sqrt{2}}c_\nu v_\phi\,, \nonumber \\
&&m_{\nu_4} \approx \frac{1}{\sqrt{2}}c_\ell v_\phi\,.
\end{eqnarray}
In this limit, the eigenstate $\nu_4$ is mostly composed of the electroweak  doublet neutrinos 
$\nu^{\prime\prime}_R$ and $\nu^\prime_L$, while $\nu_X$ is a combination of the singlets 
$\nu^{\prime\prime}_L$ and $\nu^\prime_R$ and with tiny couplings to the SM $W^\pm$ and $Z$ bosons. Since the doublet 
neutrino $\nu_4$ couples directly to the $Z$ boson, direct detection experiments render it  unacceptable as a DM 
candidate. Therefore we require $c_\nu < c_\ell$, such that $\nu_X$ is the DM candidate. Of course $\nu_4$ must be able 
to decay which means that at least one of the Yukawa couplings $y^{\prime}_\nu, y^{\prime\prime}_\nu$ should be 
nonzero to allow $\nu_4$ to decay into a Higgs boson and $\nu_X$.  Nonetheless, 
this requirement allows the $y_\nu$'s to be small enough
so as to be completely irrelevant in the discussion below.

\subsection{Annihilation Channels}
\label{sec:annihilation}

In~\cite{Joglekar:2012vc}, annihilation through the interactions generated by $y^{\prime}_\nu, y^{\prime\prime}_\nu$ 
was shown to give the correct relic abundance for DM with dominantly Majorana masses $\lesssim 100$ GeV. 
Here, because direct detection constraints require 
$y^{\prime}_\nu, y^{\prime\prime}_\nu$ to be tiny, one would have to either
rely on co-annihilation with one of the charged leptons or annihilation through a nearly on-shell Higgs. 
We instead will assume in the following that these couplings are too tiny to affect the DM phenomenology directly.

\begin{figure}
\includegraphics[width=0.35\textwidth]{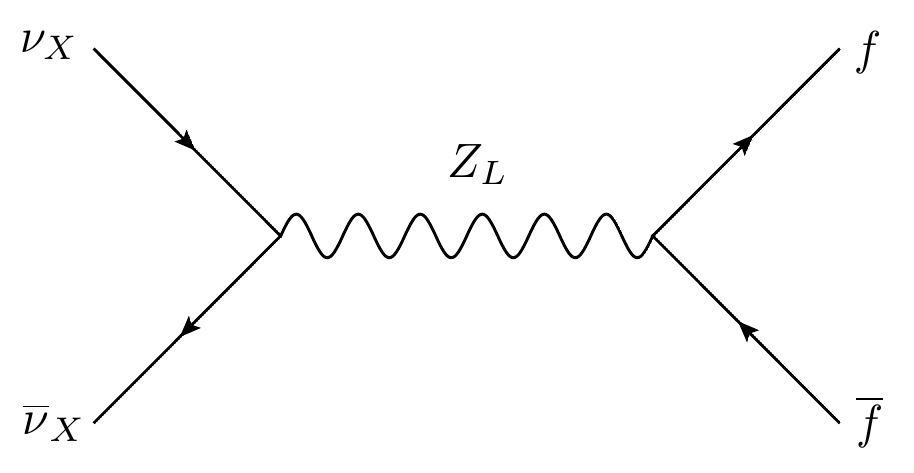}
\includegraphics[width=0.35\textwidth]{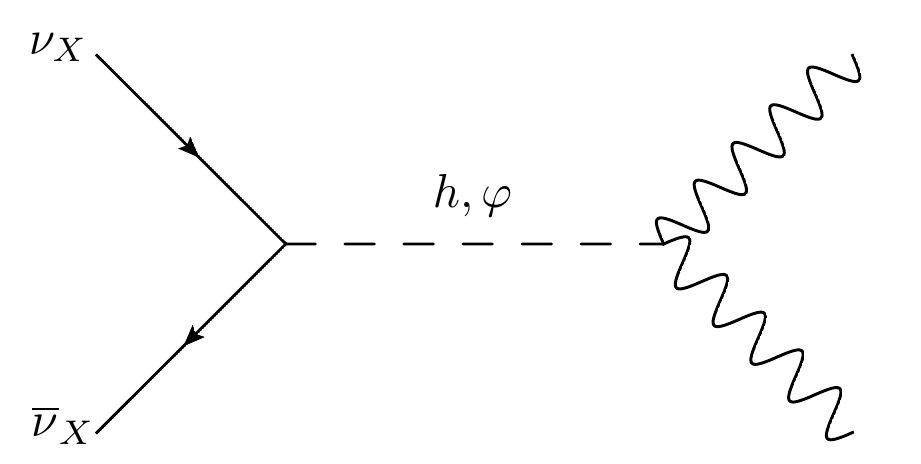}
\caption{Diagrams leading to $s$-channel $\nu_X \overline{\nu}_X$ annihilation into SM states through
exchange of $Z_L$, $h$, or $\phi$. 
\label{fig:Zpannh}}
\end{figure}

Compared to~\cite{Joglekar:2012vc}, there are additional annihilation channels for $\nu_X$ into SM leptons. 
In particular, since $\nu_X$ is a Dirac fermion, annihilation through a vector boson is $s$-wave and unsuppressed,
in contrast to the case of Majorana DM.  Indeed, the left- and right-handed components of $\nu_X$ carry lepton 
number $L''$ and $L'$, respectively, and 
$L'-L''=-3$ implies a non-vanishing coupling of $\nu_X$ to $Z_L$,
allowing $\nu_X \bar{\nu}_X$ to annihilate into SM leptons through $s$-channel 
$Z_L$ exchange, shown in the top diagram of Fig.~\ref{fig:Zpannh}. 
There are
additional annihilation channels which arise through mixing in the neutrino as well as in the Higgs sectors. 
We discuss the various annihilation modes in more detail below, assuming that $\nu_X$ is mostly singlet with at most a small 
doublet component, i.e. $y^{\prime}_\nu v_h, y^{\prime\prime}_\nu v_h \ll c_{\nu} v_\phi$. 

If $\nu_X$ acquires a small doublet component through nonzero $y^{\prime}_\nu, y^{\prime\prime}_\nu$ couplings, 
annihilation into SM particles through $Z$ or $h$ exchange becomes possible, but again we will assume that these couplings 
are sufficiently small such that these annihilation channels can be neglected. This is also required since otherwise a large direct 
detection cross section through $Z$ boson exchange would be induced. 
At the same time this suppresses annihilation into $W^+W^-$ through a heavy charged lepton exchanged
in the $t$-channel. 

The dark matter also couples to the singlet scalar $\phi_o$ with a strength $c_\nu \approx \sqrt{2} m_{\nu_X}/v_\phi$. 
When the Higgs mixing angle $\theta$ is nonzero this will allow annihilation into SM particles through $s$-channel 
exchange of $h$ and $\phi$, shown in the bottom diagram of Fig.~\ref{fig:Zpannh}. While not generally negligible, 
the contribution of these annihilation channels turns out to be suppressed compared to the $Z_L$ channel in the 
regime of interest where $v_\phi \sim 1.7$ TeV and DM $m_{\nu_X} \sim v_h$, leading to somewhat small values for $c_\nu$. 
Furthermore the $Z_L$ channel leads to unsuppressed annihilation into all SM leptons, while most of the scalar channels are 
suppressed by the small Yukawa couplings of the SM quarks and leptons to $h_o$ or small loop induced couplings of $\phi_o$ to the SM. We thus expect annihilation through $Z_L$ to be the 
dominant contribution to the relic abundance in this regime. Note also that in this regime we have $m_{\nu_X} \ll M_{Z_L}$ 
which as we will see leads to a relic abundance which is largely independent of the lepton gauge coupling $g^\prime$ 
(see Eq.(\ref{eqn:coeffs_smallR})).

\subsection{Relic Abundance}
\label{sec:relic_abundance}

Motivated by the requirement for small $y^\prime_{\nu}, y^{\prime \prime}_\nu$, we
first consider the dominant annihilation through the $Z_L$ into SM lepton pairs, and then demonstrate that
scalar exchange is unlikely to change the over-all picture. 
The relevant interactions come from Eq.(\ref{eqn:ZLag}) which before lepton number and electroweak symmetry 
breaking can be written as,
\begin{eqnarray}
\label{eqn:ZDMvert}
&&\mathcal{L} \supset g^\prime Z_{L\mu} \Big( L^{\prime\prime} \bar{\nu}^{\prime\prime}_R \gamma^\mu \nu^{\prime\prime}_R 
+ L^{\prime} \bar{\nu}^{\prime}_L \gamma^\mu \nu^{\prime}_L +  \bar{l} \gamma^\mu l  \Big) , 
\end{eqnarray}
where $l$ runs over SM leptons all of which have $L=1\,,$ which implies that the left and right handed couplings of the SM leptons 
to $Z_L$ are equal. This is in contrast to the case for the exotic leptons since $L^\prime \neq L^{\prime\prime}$. After lepton 
number breaking and rotating to the mass basis Eq.(\ref{eqn:ZDMvert}) becomes
\begin{eqnarray}
\label{eqn:ZDMvert2}
&&\mathcal{L} \supset g^\prime Z_{L\mu} \Big(\bar{\nu}_X \gamma^\mu 
(L^{\prime\prime} P_R + L^\prime P_L) \nu_X +  \bar{l} \gamma^\mu l  \Big) , 
\end{eqnarray}
 where $P_R$ and $P_L$ are the right and left projection operators respectively and we have neglected any mixing 
 between $\nu_X$ and $\nu_4$ generated by $y^{\prime}_\nu, y^{\prime\prime}_\nu$. Using Eq.~(\ref{eqn:ZDMvert2}) a 
 straight forward calculation of the diagram in Fig.~\ref{fig:Zpannh} gives the annihilation cross section,
\begin{eqnarray}
\label{eqn:annhilation_cxn}
\hspace*{-0.35cm}
\sigma = \frac{g^{\prime 4} ( (L^{\prime 2} + L^{\prime\prime 2})(s- m^2_{\nu_X}) + 6L^\prime L^{\prime\prime} m_{\nu_X}^2 ) }
{8\pi(1-4m^2_{\nu_X}/s)^{1/2}((M^2_{Z_L}-s)^2 + M^2_{Z_L}\Gamma_{Z_L}^2)},
\end{eqnarray}
where an overall factor of 6 is implicit for the three generations of charged leptons and neutrinos in the SM. 
As is well known, the annihilation cross section $\langle\sigma v\rangle$ is well approximated by a non-relativistic expansion,
$s = 4m_{\nu_X}^2 + m_{\nu_X}^2 v^2,$ and expanding the annihilation cross section in powers of $v$ to 
give $\langle \sigma v \rangle = a + b \langle v^2 \rangle + \mathcal{O}(\langle v^4 \rangle)$~\cite{Servant:2002aq}. Expanding 
Eq.~(\ref{eqn:annhilation_cxn}) we obtain
\begin{eqnarray}
\label{eqn:coeff_a}
a = \frac{3 g^{\prime 4} R^4 (L^{\prime} + L^{\prime\prime})^2}{4\pi m^2_{\nu_X}(1-4R^2)^2}
\end{eqnarray}
for the velocity independent coefficient and have defined $R = m_{\nu_X}/M_{Z_L}$ while neglecting terms of order 
$\Gamma_{Z_L}/M_{Z_L}$. 
Note, this is in contrast to the case of Majorana dark matter annihilating through a gauge boson, in which case $a=0$ up to 
corrections that are suppressed by the final state fermion masses. For the $\langle v^2 \rangle$ coefficient we have
\begin{eqnarray}
\label{eqn:coeff_b}
b = \frac{g^{\prime 4} R^4 
\Big( (L^{\prime 2} + L^{\prime\prime 2})(11 + 4R^2) + L^{\prime} L^{\prime\prime} (6 + 72R^2) \Big)
}{32 \pi m^2_{\nu_X} (1- 4R^2)^{3}}\nonumber\,. \\
\end{eqnarray}
In general the contribution from $a$ will dominate since the contribution from $b$ is suppressed by the relatively small
value of $v^2$ at freeze-out. 
It is useful to consider the limit of heavy $Z_L$ mass compared to the DM mass, or $R \ll 1$. 
Keeping only the leading term after expanding in powers of $R$ we have 
\begin{eqnarray}
\label{eqn:coeffs_smallR}
a \approx \frac{3 g^{\prime 4} (L^{\prime} + L^{\prime\prime})^2R^4}{4\pi m^2_{\nu_X}} + \mathcal{O}(R^6)
\end{eqnarray}
Since $M_{Z_L} = 3 g' v_\phi$, the dependence on the gauge coupling $g'$ cancels in the leading term,
as is usual for the contact interaction that describes vector exchange at low energies. 
For a fixed choice of the quantum numbers $L'$ and $L''$, the annihilation rate is therefore largely determined by the 
ratio $m_{\nu_X}^2/v_\phi^4$. 

\begin{figure}
\includegraphics[width=0.46\textwidth]{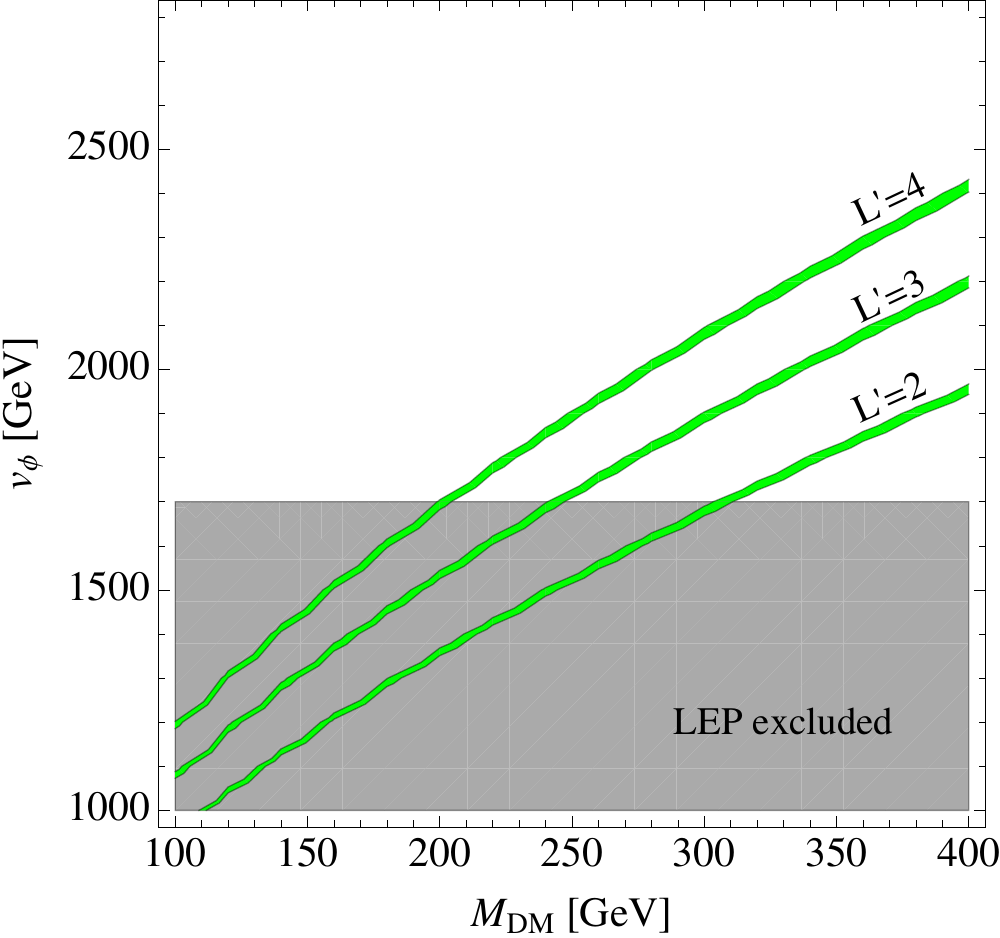}
\caption{Relic density as a function of DM mass and {\it vev} $v_\phi$, in the absence of mixing and taking $M_{DM} = m_{\nu_X}$. The 
green bands indicate regions in agreement with the measured value of 
$\Omega h^2 = 0.120\pm 0.003$~\cite{Ade:2013lta} for different choices of $L'$, as indicated in the figure.}
\label{fig:rd1}
\end{figure}

From these results a good approximation for the relic density can be obtained e.g. using the procedure 
presented in~\cite{Servant:2002aq}. We have opted instead to implement the model into the numerical code 
MICROMEGAS~\cite{Belanger:2010gh}. Not only does this facilitate the exploration of regions of parameter space where 
the ${\cal O}(v^2)$ expansion breaks down, but it also simplifies the computation of direct and indirect detection signals. 
The approximate calculation of the relic density following~\cite{Servant:2002aq} was used as validation of the 
MICROMEGAS implementation of the model.  The resulting relic density (including all sub-leading effects)
is shown as a function of $m_{\nu_X}$ and $v_\phi$, for a few choices of $L^\prime$, in Figure~\ref{fig:rd1}.
The LEP II constraints on $v_\phi$ require dark matter masses greater than about 200 GeV, and (depending on
$L^\prime$), a thermal relic density enforces a tight correlation between $v_\phi$ and $m_{\nu_X}$.

In the limit $y^{\prime}_\nu, y^{\prime\prime}_\nu \approx 0$, DM couples to $h$ and $\phi$ through $c_\nu$
and the Higgs mixing,
\begin{eqnarray}
\label{eqn:DM_scalar_couplings}
\mathcal{L} \supset \frac{c_\nu}{\sqrt{2}} ( c_\theta \phi - s_\theta h) \bar{\nu}_X \nu_X ,
\end{eqnarray}
where we have used Eq. (\ref{eqn:s1s2_definition}). These couplings allow the DM to annihilate through the bottom diagram 
shown in Fig.~\ref{fig:Zpannh}. Since dark matter masses of order the weak scale require a relatively small $c_\nu$, annihilation 
through Higgs exchange only has a small effect on the relic density. On the other hand it is crucial for direct detection which will 
be discussed in the next section.

\subsection{Direct and Indirect Detection}
\label{sec:DMpheno}

In the limit $y^{\prime}_\nu, y^{\prime\prime}_\nu \rightarrow 0$ and negligible mixing in the Higgs sector, 
the dark matter couples to SM
leptons through $Z_L$, but has no tree level interactions with quarks.  This is a challenging situation
for dark matter direct detection experiments, because of the wave function suppression to scatter off of
atomic electrons or loop suppression of the induced dark matter dipole moment \cite{Fox:2011fx}.
Consequently, even a small amount of $Z-Z_L$ or $H-\Phi$ mixing can dominate the rate, which effectively
disconnects the expectations at direct detection experiments from the relic density. 

\begin{figure}
\includegraphics[width=0.45\textwidth]{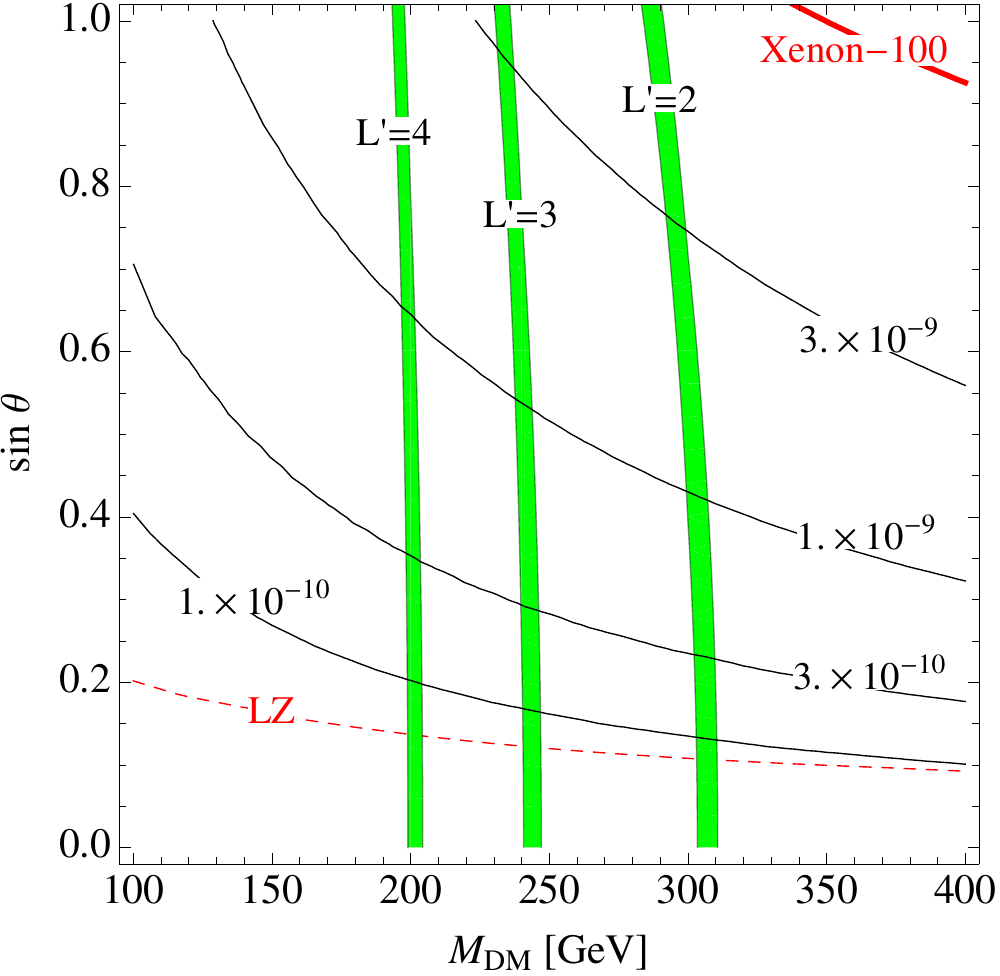}
\caption{DM-nucleon cross section in pb, as a function of the Higgs mixing angle $\sin\theta$ and of the DM mass, for $v_\phi=1.7$~TeV and $M_{DM} = m_{\nu_X}$. The solid red line indicates the current limit from the Xenon-100 experiment~\cite{Aprile:2012nq}, while the dashed red line indicates the projected reach of the LZ experiment~\cite{Malling:2011va}. The green bands indicate regions with correct relic density for different values of $L'$. }
\label{fig:DMdirect}
\end{figure}

Higgs exchange leads to spin-independent scattering with nuclei.  We compute the rate as a function of the DM mass and Higgs 
mixing angle $\sin\theta$ using MICROMEGAS and present the results in Figure~\ref{fig:DMdirect} for DM 
masses $100 - 400$~GeV.  For moderate Higgs mixing, the DM-nucleon cross section lies about one order of magnitude 
below the current best limit from the XENON-100 experiment, but is well in reach of second 
generation DM direct detection experiments such as LZ~\cite{Malling:2011va}. 

\begin{figure}
\includegraphics[width=0.215\textwidth]{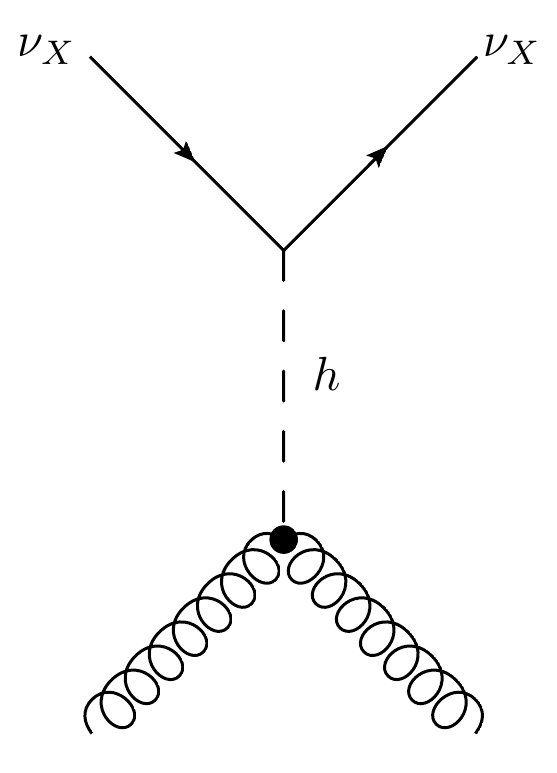}
\includegraphics[width=0.23\textwidth]{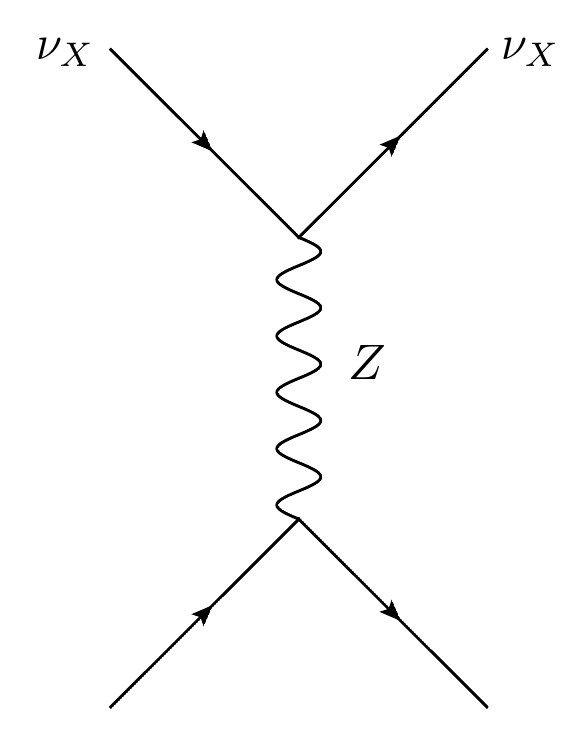}
\caption{Diagrams leading to scattering with nucleons mediated by exchange of a Higgs or $Z$ boson.}
\label{fig:DMdiagram}
\end{figure}

\begin{figure}
\includegraphics[width=0.45\textwidth]{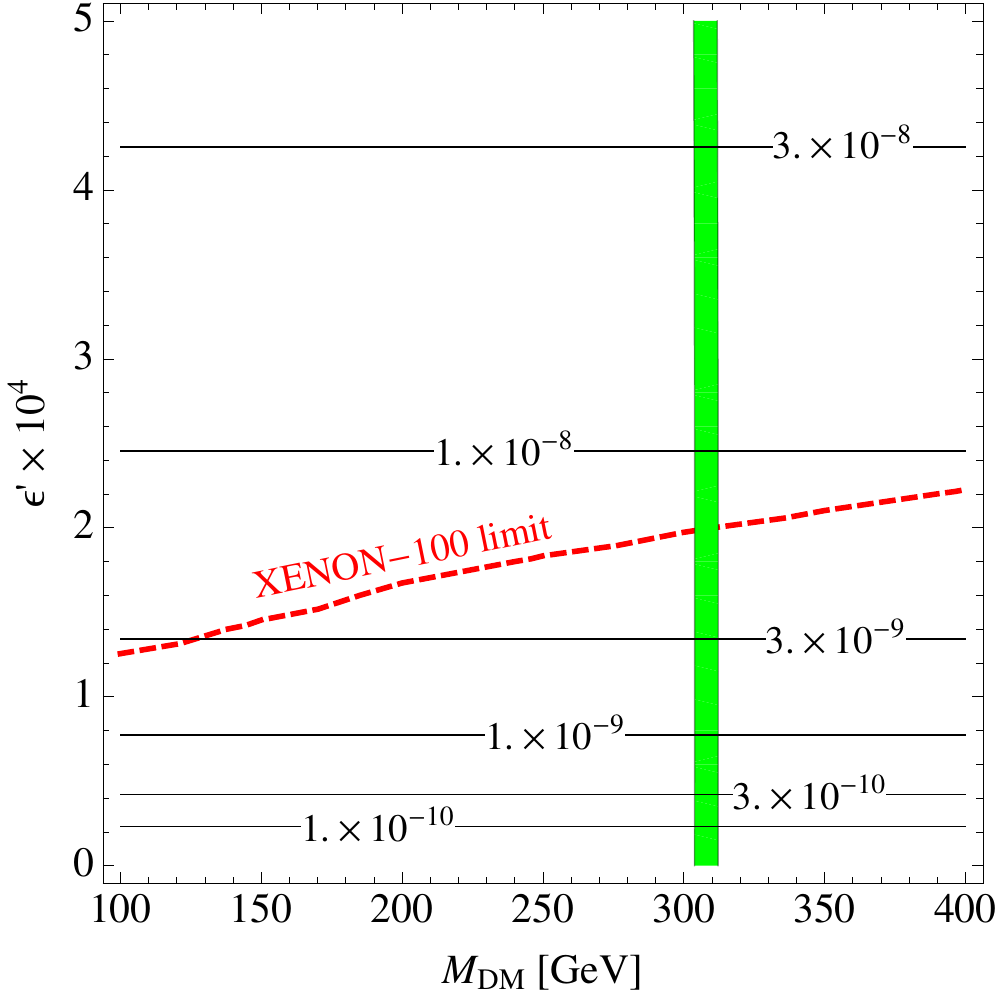}
\caption{DM-nucleon cross section in pb, as a function of the DM-Z coupling parameter $\epsilon'$ and of the DM mass (where $M_{DM} = m_{\nu_X}$), for $v_\phi=1.7$~TeV, $L'=2$ and $g'=0.5$ which implies $M_{Z_L} = 2.55$~TeV. The red dashed line indicates the current limit from the XENON-100 experiment. }
\label{fig:DMdirect2}
\end{figure}

$Z$-boson exchange induces a large DM-neutron cross section due to the sizable coupling of the $Z$
to light quarks. We parameterize the coupling of the $Z$-boson to the DM as,
\begin{eqnarray}
\label{eqn:ZLDMcoupling}
	{\cal L} \supset \epsilon' g' Z_\mu  \bar\nu_X \gamma^\mu  \left( L'' P_R + L' P_L \right) \nu_X  \,,
\end{eqnarray}
where $\epsilon'$ is either induced by $Z-Z'$ mixing or by nonzero neutrino Yukawa couplings 
$y^{\prime}_\nu, y^{\prime\prime}_\nu$. The upper bound on $\epsilon'$ from direct detection for $L'=2$ 
is shown in Fig.~\ref{fig:DMdirect2}, for DM masses $100 - 400$~GeV.  One can see that  for $g^\prime = 0.5$ 
and $v_\phi = 1.7$~TeV, direct detection requires roughly $\epsilon^\prime \lesssim 1-2\times 10^{-4}$ depending 
on the DM mass. In the limit $y^{\prime}_\nu, y^{\prime\prime}_\nu \approx 0$, $\epsilon^\prime$ is due 
solely to $Z-Z_L$ mixing and gives $\epsilon'=\sin\xi$ as defined in Eq.(\ref{eqn:Zmixingangle}). 
Since $M_{Z_L} = 3 g^\prime v_\phi = 2.55$~TeV, Eq.(\ref{eqn:Zmixingangle}) and Fig.\ref{fig:ep_vs_mZL} 
together imply that for a gauge kinetic mixing parameter (see Eq.(\ref{eqn:ZLag})) of $\epsilon \sim 7\times10^{-3}$ direct detection signals roughly 20 times below the current bound can be obtained, within range of future direct detection experiments~\cite{Malling:2011va}.

Dark matter can also be observed indirectly, by searching for the products of DM annihilation.  Here, the dark matter annihilates 
predominantly into charged leptons or neutrinos.  While there is a large rate into positrons, it is characterized by roughly the 
thermal relic cross section and is thus quite a bit too small to account for the
anomalous positron fraction observed by PAMELA~\cite{Adriani:2008zr}, Fermi~\cite{FermiLAT:2011ab}, and 
AMS-02~\cite{Aguilar:2013qda}.  At the same time, contributions to the anti-proton flux are very tiny,
evading constraints from PAMELA~\cite{Adriani:2010rc}.

Annihilation into charged leptons will also produce gamma rays as secondaries.  Currently, the tightest constraints on such 
production are from the Fermi LAT null observations of dwarf spheroidal galaxies~\cite{Ackermann:2011wa}, which are just 
short of being able to  rule out thermal cross sections for dark matter masses around a few 10's of GeV based on one sixth of 
the annihilations producing $\tau^+ \tau^-$.  In the near future, such constraints are only relevant for $\nu_X$ dark matter which 
has been produced non-thermally.

Dark matter may also annihilate directly into $\gamma \gamma$ and/or $\gamma Z$ at loop level, providing mono-chromatic
gamma ray lines, whose distinctive energy profile can help compensate for a tiny rate.  Predictions for the class of models
including $U(1)_L$ were studied in \cite{Jackson:2013pjq}, where it was found that $\gamma \gamma$, $\gamma Z$, and
$\gamma \phi$ (if kinematically accessible) final states can be generated. The largest signal is likely to be $\gamma \phi$, which is
expected to be at least an order of magnitude below the current Fermi bounds \cite{Ackermann:2012qk},
but may be visible to future experiments.

The rate for dark matter to be captured in the Sun or Earth and then annihilate into high energy neutrinos  is controlled by the 
spin-dependent cross section which in turn is controlled by the degree of $Z-Z_L$ mixing. Thus, despite a large annihilation 
fraction into SM neutrinos, the precision constraints render it difficult to imagine an observable rate at ICECUBE in the near 
future~\cite{Aartsen:2012kia}.


\section{LHC Phenomenology and Constraints}
\label{sec:lhc_pheno}

The presence of new particles required by the $U(1)_L$ gauge symmetry leads to a variety of potentially interesting LHC 
phenomenology. In this section we discuss various aspects of the phenomenology of this model as well as the relevant 
constraints coming from the LHC. We also examine in more detail the charged lepton sector and its effects on the Higgs decays.

\subsection{Exotic Charged Lepton Sector}
\label{sec:charged_leptons}
Once $\Phi$ and $H$ obtain expectation values, the Lagrangian for the exotic charged lepton sector becomes,
\begin{eqnarray}
\label{eqn:eyuklag}
&&\mathcal{L} \supset -\frac{c_\ell v_\phi}{\sqrt{2}} (1 + \frac{\phi_o}{v_\phi}) \bar{e}^{\prime\prime}_{R}e^{\prime}_{L} 
-\frac{c_e v_\phi} {\sqrt{2}} (1 + \frac{\phi_o}{v_\phi})\bar{e}^{\prime\prime}_{L}e^{\prime}_{R} \\
&& -\frac{y^{\prime\prime}_e v_h}{\sqrt{2}} (1 + \frac{h_o}{v_h}) \bar{e}^{\prime\prime}_{R}e^{\prime\prime}_{L} 
- \frac{y^{\prime}_e v_h}{\sqrt{2}} (1 + \frac{h_o}{v_h}) \bar{e}^{\prime}_{L} e^{\prime}_{R} + \emph{h.c.}\nonumber 
\end{eqnarray}
which gives a mass matrix of the same form as that found in the neutrino sector,
\begin{equation}
\label{eqn:e_mass_matrix}
\mathcal{M}_e = \frac{1}{\sqrt{2}} 
\left( \begin{array}{cc}
 c_\ell v_\phi & y^{\prime\prime}_e v_h \\
 y^{\prime}_e v_h &  c_e v_\phi \\
\end{array} \right) \ .
\end{equation}
Again we can diagonalize via $\mathcal{M}_{e D} = U^\dagger_L \mathcal{M}_e U_R$ to obtain the mass eigenvalues and 
eigenstates. The Lagrangian in Eq.(\ref{eqn:eyuklag}) also leads to the interaction matrices for $\phi_o$ and $h_o$ given by,
\begin{equation}
\label{eqn:e_int_matrix}
\mathcal{N}^h_e = \frac{v_h}{\sqrt{2}} 
\left( \begin{array}{cc}
 0 & y^{\prime\prime}_e \\
 y^{\prime}_e &  0 \\
\end{array} \right), 
~
\mathcal{N}^\phi_e = \frac{v_\phi}{\sqrt{2}} 
\left( \begin{array}{cc}
 c_\ell  & 0 \\
 0 &  c_e \\
\end{array} \right),
\end{equation}
which upon the rotation performed to diagonalize $\mathcal{M}_e$ gives interaction matrices in the mass basis defined as 
$\mathcal{V}_\phi=U^\dagger_L  \mathcal{N}^\phi_e U_R$ and $\mathcal{V}_h=U^\dagger_L  \mathcal{N}^h_e U_R$. 
These matrices dictate the couplings of the exotic leptons to $\phi$ and $h$. We note also that Eq.(\ref{eqn:e_mass_matrix}) is 
the same mass matrix in the charged lepton sector considered in~\cite{Joglekar:2012vc}, with the difference being that in this 
model there are no explicit mass terms. In particular, when $v_{h}, v_\phi\rightarrow 0$ all masses go to zero, which
makes the gauged lepton number model more constrained and relates the electroweak and lepton 
breaking scales to the rate of Higgs decay to di-photons, as we will see below.

A useful simplifying limit is $c_\ell \approx c_e \equiv c_e$ and $y^\prime_e \approx y^{\prime\prime}_e \equiv y_e$  
in which case the charged leptons are maximally mixed and one obtains the simple relations for the mass eigenvalues,
\begin{eqnarray}
\label{eqn:emasses}
&&m_{e_1} \approx \frac{1}{\sqrt{2}}(c_ev_\phi - y_ev_h)\nonumber \\
&&m_{e_2} \approx \frac{1}{\sqrt{2}}(c_ev_\phi + y_ev_h)\,,
\end{eqnarray}
where we have assumed $c_e v_\phi > y_ev_h$. Thus we see that for fixed $y_e$ and $v_\phi$, the mass of the charged 
leptons is controlled by $c_e$. Along with the scalar mixing discussed in Sec.~\ref{sec:gauge_higgs_sector} we now have the 
pieces necessary for examining the modification to Higgs decays.

\subsection{Modifications of Higgs Decays}
\label{sec:diphoton_rate}

Assuming that the Higgs can not decay directly into new particles, the primary effect of the new lepton sector on 
Higgs decays will be through loop effects. From the discussion on Higgs mixing in Sec.\ref{sec:gauge_higgs_sector}, 
we can write the modification of the SM Higgs partial width as,
\begin{eqnarray}
\label{eqn:epsilon_definition}
&& \epsilon_{i} \equiv  \dfrac{\Gamma_{h i}}{\Gamma^{SM}_{h_o i}}
 =  \dfrac{ \left| \mathcal{M} (h \rightarrow i)
  \right|^2 }{ \left| \mathcal{M} (h_o \xrightarrow[SM]{} i) \right|^2 } \nonumber \\
&& = \dfrac{  c^2_\theta\left|
      \mathcal{M} (h_o \rightarrow i) 
     - t_\theta \mathcal{M} (\phi_o \rightarrow  i) \right|^2}{
\left| \mathcal{M} (h_o \xrightarrow[SM]{} i) \right|^2 } ,
\end{eqnarray}
where we have used Eq.(\ref{eqn:s1s2_definition}) and $\Gamma^{SM}_{h_o i}$ is the SM partial width to a final 
state $i$ and $\Gamma_{h i}$ is the partial width for $h$ to decay into $i$. 
The rate expected at the LHC relative to the SM can be written as,
\begin{equation}
\label{eqn:rate}
\mu_i = \dfrac{\sigma (j \rightarrow h)}{\sigma (j \xrightarrow[SM]{} h_o)} 
\dfrac{\mathcal{B}(h \rightarrow  i)}{\mathcal{B} (h_o \xrightarrow[SM]{} i)} 
= \epsilon_{j} \frac{\Gamma^{SM}_{h_o}}{\Gamma_h} \epsilon_{i}\,,
\end{equation}
where we have made use of the narrow width approximation, $\mathcal{B}$ signifies the branching fraction, and the production 
channels are labeled $j = VV, gg$. We also define $\Gamma^{SM}_{h_o}$ as the total SM Higgs width and $\Gamma_h$ as 
the total decay width for the mass eigenstate $h$. Since this model does not contain any new colored particles the only new 
effects entering $\epsilon_{gg}$ are through Higgs mixing which gives $\epsilon_{gg}\approx c^2_\theta$. Since $ZZ$ 
and $WW$ already occur at tree level in the SM, we assume the loop corrections due to the new leptons are negligible 
which implies the only effect again comes from Higgs mixing, which gives 
$\epsilon_{ZZ} =\epsilon_{WW} \approx c_\theta^2$. Similarly for the SM Higgs Yukawa interactions 
we have $\epsilon_Y \approx c_\theta^2$.

This leaves the $Z\gamma$ and $\gamma\gamma$ channels, which first occur at one loop in the SM, as the most 
promising possibilities for these effects to manifest themselves. However, in Refs.~\cite{Carena:2012xa,Joglekar:2012vc}
the modification to $Z\gamma$ was 
shown to be only $\approx 5\%$ for a corresponding $\gamma\gamma$ enhancement  of $\approx 50\%$,
and to good approximation $\epsilon_{Z\gamma} \approx c_\theta^2$. Thus, in addition to the 
universal $c_\theta^2$ suppression from Higgs mixing, the only additional modifications to the total decay width comes 
from the $\gamma\gamma$ channel through loops of exotic charged leptons. Since for the modifications we are 
interested in $\Gamma_{h\gamma\gamma} \ll  \Gamma_h$ this implies 
$\Gamma^{SM}_{h_o}/\Gamma_h \approx c^{-2}_\theta$ which will cancel with the $c^2_\theta$ in the production channel ratios 
$\epsilon_{gg,VV}$. 
This gives finally for the relative rates $\mu_{i} = c^2_\theta$ for $i\neq \gamma\gamma$ 
and for the final modified diphoton signal strength,
\begin{eqnarray}
\label{eqn:diphoton_rate}
\mu_{\gamma\gamma}=\epsilon_{\gamma\gamma}.
\end{eqnarray}

Using the approach and conventions of~\cite{Kumar:2012ww}, which examined the 
similar $gg\rightarrow h$ process, we can go on to obtain the exotic charged lepton 
contributions to the $h\rightarrow \gamma\gamma$ amplitude by computing $h_o \rightarrow \gamma\gamma$ and $\phi_o\rightarrow \gamma\gamma$ (omitting photon polarization vectors),
\begin{widetext}
\begin{eqnarray}
\label{eqn:amp_def}
&& \mathcal{M}^{\mu\nu}({h_o \rightarrow \gamma \gamma})  =  (\frac{\alpha}{2 \pi v_h}) 
\sum \limits_i \dfrac{(\mathcal{V}_h)_{ii}F_F(\tau_{e_i}) }{m_{e_i}}  
\Big( p_1^\nu p_2^\mu - \frac{m_h^2}{2} g^{\mu\nu} \Big) \nonumber \\
&& \mathcal{M}^{\mu\nu}({\phi_o \rightarrow \gamma \gamma})  =  (\frac{\alpha}{2 \pi v_\phi}) 
\sum \limits_i \dfrac{(\mathcal{V}_\phi)_{ii} F_F(\tau_{e_i})}{m_{e_i}}  
\Big( p_1^\nu p_2^\mu - \frac{m_h^2}{2} g^{\mu\nu} \Big) , \nonumber \\
\end{eqnarray}
\end{widetext}
where the index $i=1,2$ runs over the exotic charged lepton mass eigenstates found after diagonalizing the mass matrix in 
Eq.~(\ref{eqn:e_mass_matrix}), and $F_F$ are the fermonic loop functions with $\tau_{e_i} = m^2_{h}/4m^2_{e_i}$ 
as defined in~\cite{Kumar:2012ww}. Note that the amplitudes in Eq.(\ref{eqn:amp_def}) are evaluated 
at $m_{h_o} = m_h$ and $m_{\phi_o} = m_h$ where $m_h$ is the physical scalar mass. 

Using Eq.(\ref{eqn:epsilon_definition})-(\ref{eqn:amp_def}) we obtain,
\begin{widetext}
\begin{eqnarray}
\label{eqn:AArate1}
\begin{array}{l}
\mu_{\gamma\gamma} =
\frac{\left| 
\dfrac{c_\theta}{v_h} \left( F_{SM} + \sum \limits_i \dfrac{(\mathcal{V}_h)_{ii}}{m_{e_i}} F_F(\tau_{e_i}) \right)
- \dfrac{s_\theta}{v_\phi} \left( \sum \limits_i  \dfrac{(\mathcal{V}_\phi)_{ii}}{m_{e_i}} 
F_F(\tau_{e_i}) \right) \right|^2}{\Big|F_{SM}/v_h \Big|^2} \\
\\
= 
c^2_\theta \left| 
 \left(1+ F^{-1}_{SM} \sum \limits_i 
\dfrac{(\mathcal{V}_h)_{ii}}{m_{e_i}} F_F(\tau_{e_i})\right)
- t_\theta 
\left( F^{-1}_{SM} \dfrac{v_h}{v_\phi} 
 \sum \limits_i  \dfrac{(\mathcal{V}_\phi)_{ii}}{m_{e_i}} F_F(\tau_{e_i}) \right)
\right|^2 , \\
\end{array}
\end{eqnarray}
\end{widetext}
where $F_{SM}$ is the SM loop function which includes the dominant and negative $W^{\pm}$ boson contribution 
as well as the smaller and positive $t$-quark, which sum to give numerical value of $\approx -6.5$ for $m_{h} =125$ GeV.  
Note only the diagonal entries in the interaction matrices $(\mathcal{V}_h)_{ii}$ and $(\mathcal{V}_\phi)_{ii}$ contribute 
in the $h\rightarrow \gamma\gamma$ loop.

After the approximations leading to the masses in Eq.(\ref{eqn:emasses}), which 
give $(\mathcal{V}_\phi)_{11} = (\mathcal{V}_\phi)_{22} \approx c_e v_\phi/\sqrt{2}$ 
and $(\mathcal{V}_h)_{11} = -(\mathcal{V}_h)_{22} \approx -y_e v_h/\sqrt{2}$, we obtain (approximately) for the 
modified signal strength,
\begin{widetext}
\begin{equation}
\label{eqn:AArate2}
\mu_{\gamma\gamma} \simeq c_\theta^2 \Big|1- \frac{v_h}{\sqrt{2}F_{SM}}\Big[ y_e \Big( \frac{F_F(\tau_{e_1})}{m_{e_1}}  
-  \frac{F_F(\tau_{e_2})}{m_{e_2}} \Big) + c_e t_\theta 
\Big( \frac{F_F(\tau_{e_1})}{m_{e_1}} + \frac{F_F(\tau_{e_2})}{m_{e_2}} \Big) \Big]\Big|^2 , 
\end{equation}
\end{widetext}
where $m_{e_1,e_2}$ are given in Eq.(\ref{eqn:emasses}) and satisfy $m_{e_1} < m_{e_2}$. Remembering that 
$F_{SM} < 0$ we see in the limit $t_\theta \rightarrow 0$ we have an enhancement in the diphoton rate in the 
presence of mostly vector-like leptons entering through the $h_o$ component of $h$. This is, of course, expected from 
the low energy Higgs theorems (see e.g.~\cite{Joglekar:2012vc}). 
We see also that the contribution from Higgs mixing is constructive for $t_\theta > 0$ and destructive for $t_\theta < 0$ which 
also corresponds to the sign of the coupling $\lambda_{hp}$ in Eq.(\ref{eqn:scalarlag}). 
In the limit $y_e \rightarrow 0$ the enhancement enters entirely through Higgs mixing and thus requires large 
mixing angles and Yukawa coupling $c_e$. 
In the realistic limit $v_\phi \gg v_h$, the $e_1$ and $e_2$ become almost purely vector-like and again the contribution only 
enters through Higgs mixing via the $\phi_o$ component of $h$. 
However as $v_\phi \rightarrow \infty$ one also has $t_\theta \rightarrow 0$ and the $\phi_o$ contribution eventually 
decouples from the $h \rightarrow \gamma\gamma$ amplitude as $v_\phi$ is taken large. Eq.(\ref{eqn:AArate2}) is
in agreement with~\cite{Batell:2012zw} for the case where their explicit mass term is put to zero. 

To avoid the constraints discussed in Sec.~\ref{sec:constraints} we choose $v_\phi = 1.7$ TeV and take the lightest 
charged lepton to have mass greater than $m_{min} \sim$ 100 GeV. Measurements of the Higgs decays at the LHC 
indicate rates consistent with the SM with the possibility of a slight, though not significant, enhancement in the 
diphoton channel~\cite{Falkowski:2013dza}. Regardless this implies that these fermions must be mostly `vector-like' 
since otherwise their effects would lead to destructive interference~\cite{Joglekar:2012vc} with the SM contribution 
giving a reduced rate, which is disfavored. This allows us to write,
\begin{eqnarray}
\label{eqn:deltam}
&&m_{e_1} = \frac{c_e v_\phi - y_e v_h}{\sqrt{2}} \gtrsim m_{min},
\end{eqnarray}
which leads to a condition on the Yukawa coupling,
\begin{eqnarray}
\label{eqn:cmin}
&&  \frac{\sqrt{2} m_{min}  + y_e v_h}{v_\phi} \lesssim c_e \lesssim 4\pi.
\end{eqnarray}
where we have also indicated $4\pi$ as the perturbative upper bound. 

\begin{figure}[t]
\center
\includegraphics[width=0.45\textwidth]{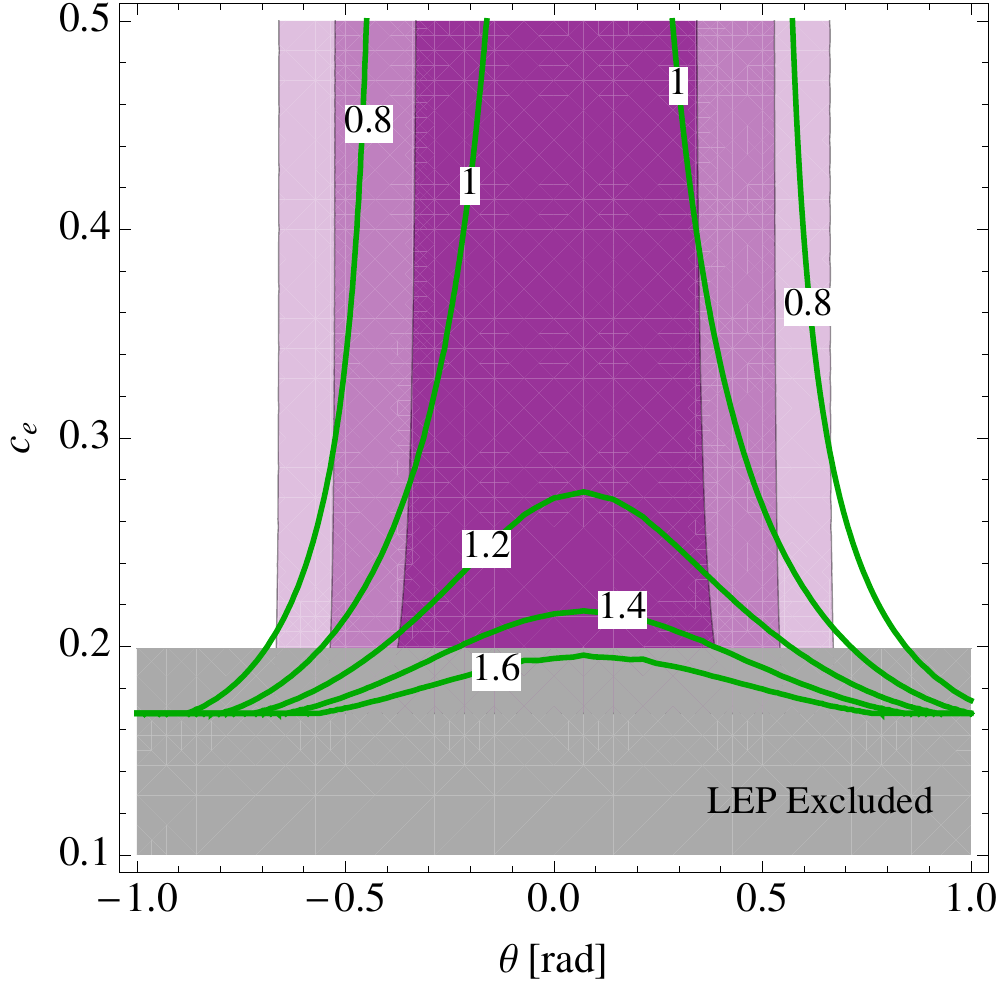}
\caption{Fits to the full Higgs data set in the $c_e - \theta$ plane for $y_e = 0.8$ and $v_\phi = 1.7$ TeV. Here the purple contours show the $1,2,3\sigma$ regions while the grey band shows the LEP excluded region the green lines are contours of constant $\mu_{\gamma\gamma}$. Details on the fitting procedure can be found in~\cite{Freitas:2012kw}.}
\label{fig:Higgs_fit}
\end{figure}

Since the mixing angle will affect all decay channels, we perform a fit to the full Higgs data~\cite{ATLAS-CONF-2013-034,CMS-PAS-HIG-13-005}
set
 in the $c_e - \theta$ plane for fixed $y_e = 0.8$ and $v_\phi = 1.7$ TeV. 
We show in Fig.~\ref{fig:Higgs_fit} the 1,~2,~3~$\sigma$ regions (purple) for the favored parameter space where the 
grey band shows the excluded region by LEP II for which $m_{e_1} < 100$ GeV. Values as large as $\theta \sim \pm 0.5$ give a good fit to the Higgs data, while larger values are disfavored due to the $\cos  \theta$ suppression of the signal rates. We also show contours of the relative diphoton rate shown in the green curves, though it is also worth noting that with the current data, the diphoton rate has no significant impact on the quality of the fit. Negative values of the mixing angle correspond to $\lambda_{hp} < 0\,,$ which can potentially lead to vacuum instabilities. On the other hand, positive values of $\theta \sim 0.5$ where $\lambda_{hp} > 0$ lead to no instability and as shown in~\cite{Batell:2012zw} can be made consistent  with constraints coming from the $S$ and $T$ parameters. 

\begin{figure}[t]
\includegraphics[width=0.45\textwidth]{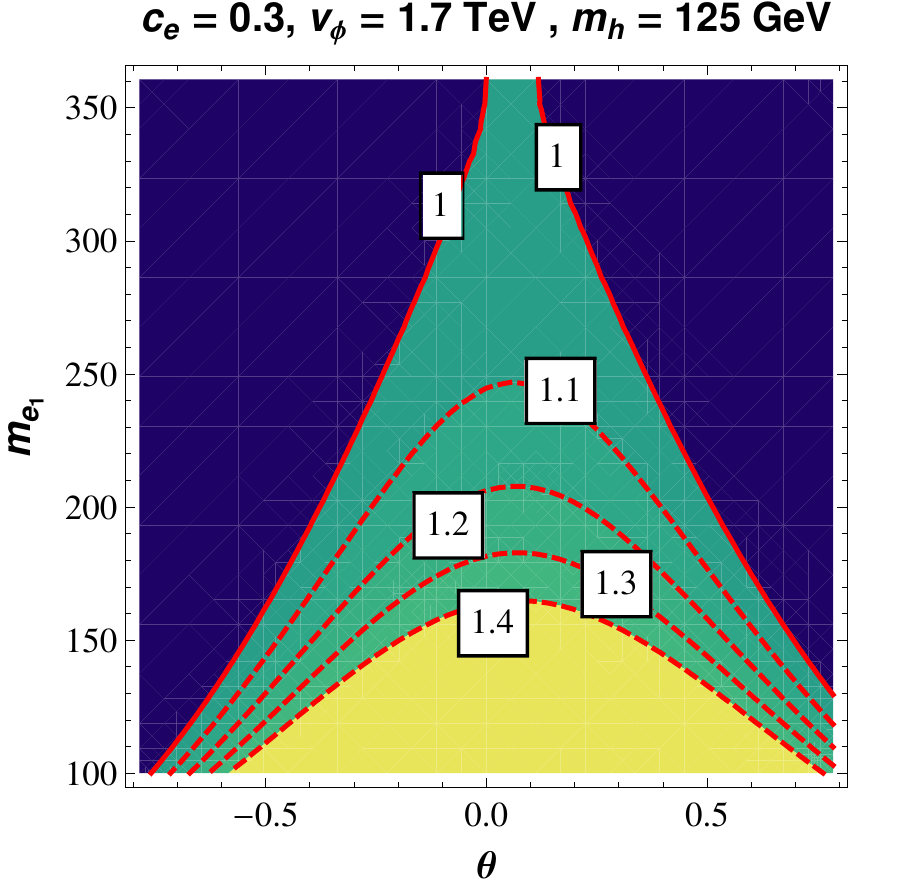}
\caption{Contours of relative diphoton rate as a function of the Higgs mixing angle $\theta$ and lightest exotic charged lepton mass $m_{e_1}$.}
\label{fig:me1_vs_th}
\end{figure}

Choosing instead to fix $c_e =0.3$ and trading in $y_e$ for the lightest charged lepton mass, we can examine contours of $\mu_{\gamma\gamma}$ as a function of $m_{e_1}$ and $\theta$ as seen in Fig.~\ref{fig:me1_vs_th}. Since the DM mass serves as a lower bound on the charged lepton mass we see for the DM masses $\gtrsim 200$~GeV found in Sec.{\ref{sec:darkmatter}} that modifications up to $\sim 10-20\%$ can be obtained for $\theta \sim 0.3 - 0.4$ and $m_{e_1} \gtrsim 200$~GeV. Of course one can lower this bound by considering larger values of $L^\prime$ as can be seen in Fig.~\ref{fig:rd1}, or by tuning the $Z_L$ mass such that the DM annihilation is resonantly enhanced.


\begin{figure}[t]
\includegraphics[width=0.45\textwidth]{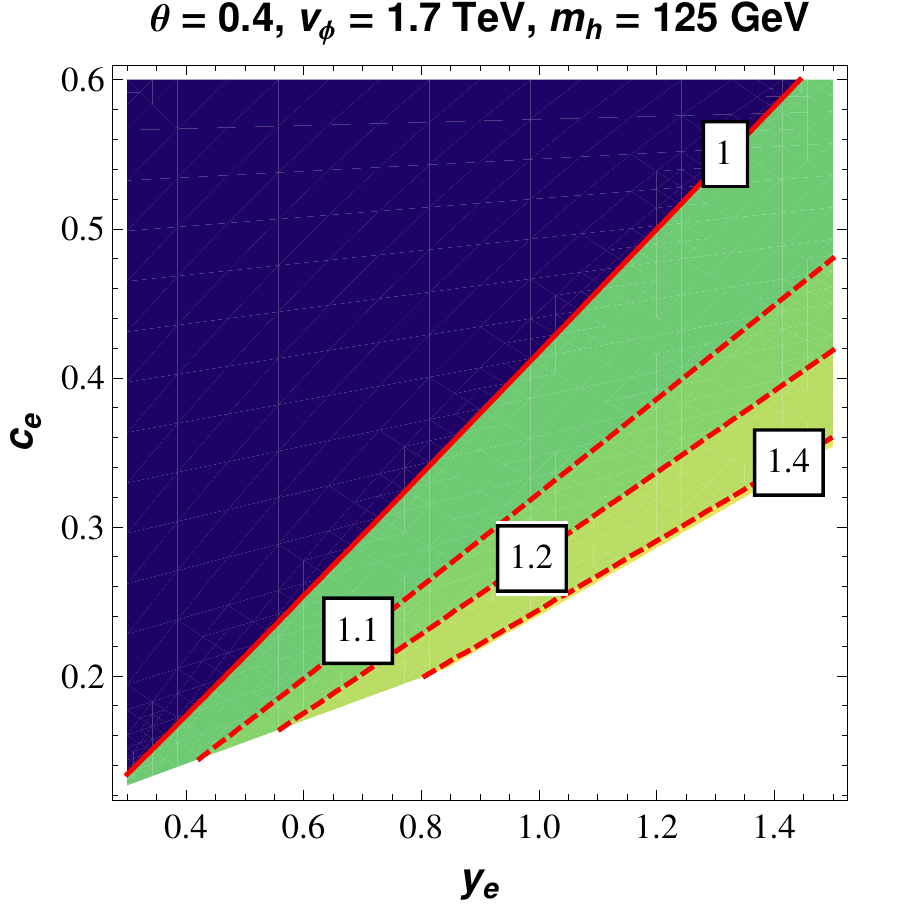}
\caption{Contours of relative diphoton rate as a function of exotic charged lepton Yukawa couplings.}
\label{fig:c_vs_ynm}
\end{figure}

Allowing $c_e$ and $y_e$ to vary instead while fixing $\theta = 0.4$ and $v_\phi = 1.7$ TeV, 
we show $\mu_{\gamma\gamma}$ contours in the $c_e - y_e$ plane in Fig.~\ref{fig:c_vs_ynm}. 
As can be seen, observable modifications can be obtained for $\mathcal{O}(1)$ values of the Yukawa couplings 
for which vacuum stability issues can be avoided~\cite{Batell:2012zw}. For these ranges of Yukawa couplings, 
$m_{e_1}$ lies in the range $100-500$ GeV, such that the exotic leptons can be produced at the LHC. We will discuss possible collider signatures below. 

\begin{figure}[t]
\includegraphics[width=0.45\textwidth]{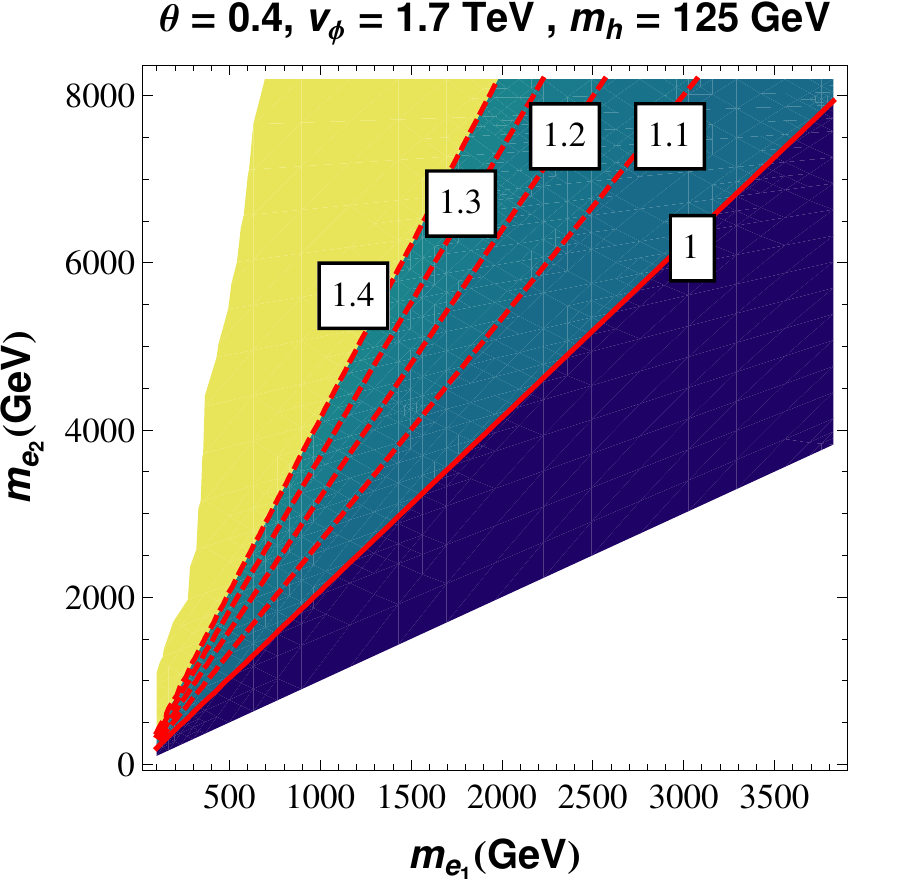}
\caption{Contours of relative diphoton rate as a function of exotic charged lepton masses. Here we allow the masses to be as large as allowed by perturbativity and $v_\phi = 1.7$~TeV.}
\label{fig:me1_vs_me2}
\end{figure}

If one is willing to push the Yukawa couplings as large as the perturbative limit $\sim 4\pi$, one can realize
large deviations in $\mu_{\gamma\gamma}$ even for multi-TeV masses.  In Figure~\ref{fig:me1_vs_me2}, we
show the deviation in the plane of $m_{e_1}$-$m_{e_2}$ for fixed $v_\phi = 1.7$ TeV, right above the LEP II limit.
Even for a lightest exotic charged lepton with mass $m_{e_1} \sim 2 - 3$ TeV, one can obtain appreciable modifications to the
Higgs diphoton rate, reflecting the fact that the fermion masses here are purely the result of Yukawa couplings,
and thus do not exhibit decoupling \cite{Appelquist:1974tg}.  Of course, all exotic contributions to the
$h\rightarrow \gamma\gamma$ amplitude decouple in the limit of $v_\phi \rightarrow \infty$. It should also be noted that the required large Yukawa couplings can induce vacuum instabilities in the Higgs potential at scales close to the masses of the exotic leptons. Additional structures like supersymmetry would be required to restore vacuum stability. Some work in this direction recently appeared in~\cite{Feng:2013mea,Joglekar:2013zya,Kyae:2013hda}.

\subsection{Other Potential LHC Signatures}
\label{subsec:LHCpheno}

Since the LHC is a hadron machine, weakly coupled extensions of the SM such as the model 
presented here are not heavily constrained by the current LHC data. Currently, constraints on the 
masses of the new leptons and of $Z_L$ mostly derive from the LEP experiments. Exotic charged 
leptons must be heavier than about 100~GeV for consistency 
with direct search limits. The $Z_L$ mass should be larger than the LEP-2 center-of-mass energy of 209~GeV, and furthermore its coupling s subject to the constraint 
$M_{Z_L} = 3 g' v_\phi$ where $v_\phi \geq 1.7$~TeV
(and we have neglected any kinetic mixing with the $Z$ boson). 

\begin{figure}[t]
\includegraphics[width=0.45\textwidth]{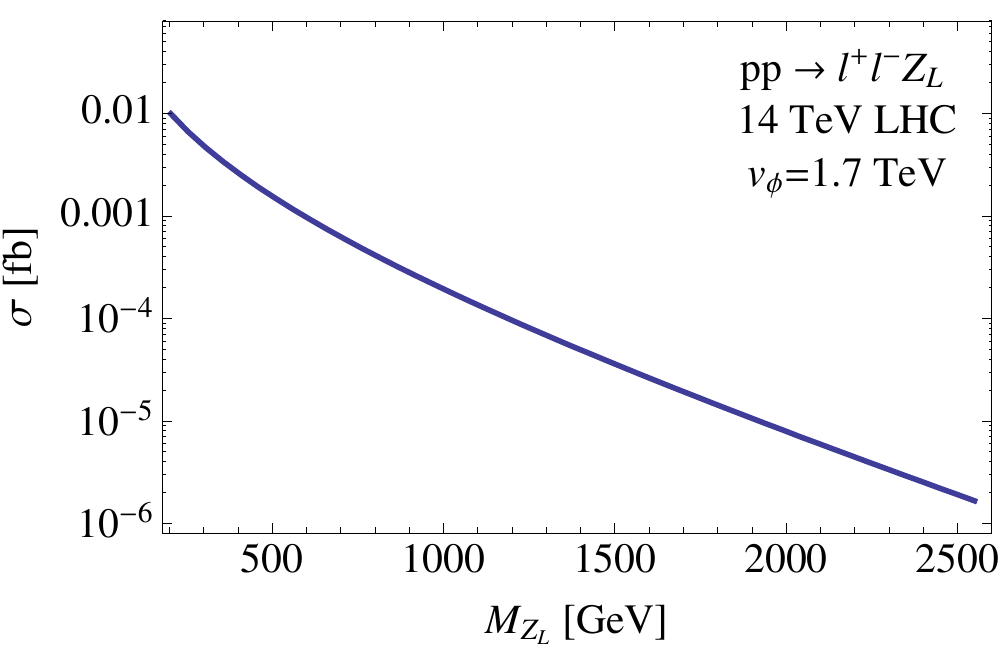}
\caption{Cross section for the process $pp \to \ell^+ \ell^- Z_L$ at the 14~TeV LHC, for $v_\phi=1.7$~TeV, and summed over SM leptons, $\ell^\pm = e^\pm, \mu^\pm, \tau^\pm$. }
\label{fig:ZLstrahlung}
\end{figure}

One of the defining features of our model is $Z_L$, the gauge boson of the lepton number symmetry. Since it does not couple to quarks, it is difficult to produce at the LHC. The most promising option is to radiate a $Z_L$ from a pair of Drell-Yan produced leptons, in the process $p p \to \ell^+ \ell^- Z_L$. The cross section for this process is calculated using the program CALCHEP~\cite{Belyaev:2012qa} with the MRST2002 PDF set~\cite{Martin:2002aw} and shown in Fig.~\ref{fig:ZLstrahlung}, where one can see it is at most of order $10^{-2}$~fb at the 14~TeV LHC. As long as the new leptons are heavier than half the $Z_L$ mass, the gauge boson will decay into charged SM leptons with a branching ratio of 50\%, while the other 50\% are into neutrinos (recalling there are three light $\nu_{Ri}$ in this model). The final state with four charged leptons, two of which reconstruct the $Z_L$ mass, is essentially background free. Nevertheless even at a possible high luminosity upgrade of the LHC with 3~ab$^{-1}$ it will be difficult to probe $Z_L$ masses above 500~GeV. 
\begin{figure}[t]
\includegraphics[width=0.45\textwidth]{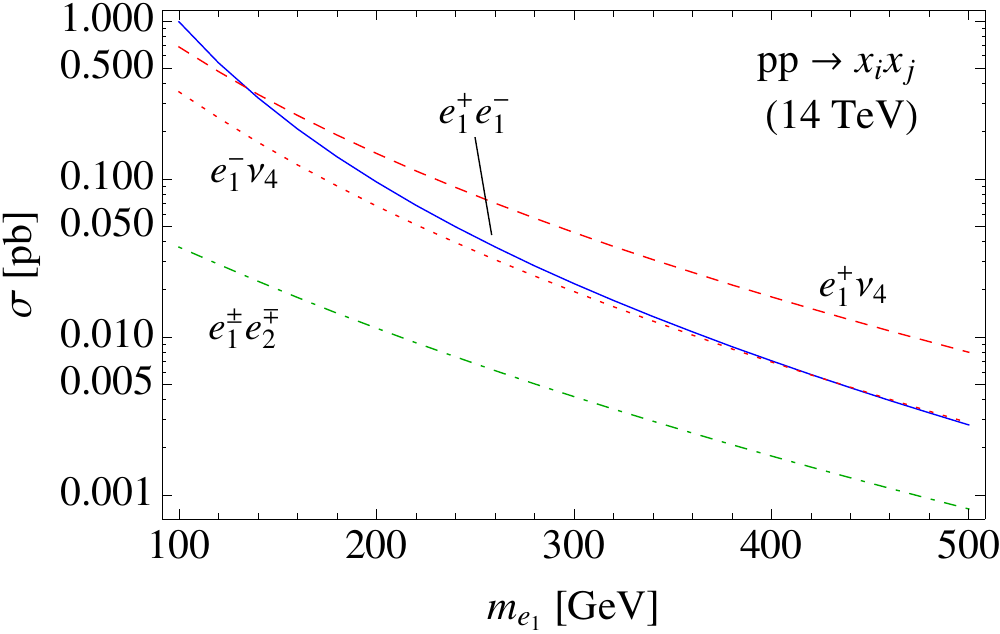}
\caption{Cross sections for the pair production of exotic leptons at the 14~TeV LHC, as a function of the lightest charged lepton mass $m_{e_1}$ in the limit leading to Eqs.(\ref{eqn:DM_masses}) and (\ref{eqn:emasses}). For the processes involving $e_2^\pm$ and $\nu_4$ we have assumed that $m_{e_2} = m_{e_1} + 280$~GeV which implies $m_{\nu_4} = m_{e_1} + 140$~GeV. }
\label{fig:LHCpair}
\end{figure}

Pairs of charged and neutal leptons can be pair produced at the LHC in the Drell-Yan process. The cross sections for the different processes at the 14~TeV LHC are shown in Fig.\ref{fig:LHCpair}, and were again obtained using CALCHEP. The processes are similar to chargino/neutralino pair production, for which NLO corrections are moderate~\cite{Beenakker:1996ed}.
For this plot we have assumed that the lepton masses are given by Eq.~(\ref{eqn:DM_masses}) and Eq.~(\ref{eqn:emasses}). This leads to the following mass hierarchies for the exotic lepton sector,
\begin{eqnarray}
\label{eqn:hier}
m_{e_2} > m_{\nu_4} > m_{e_1} > m_{\nu_X}.
\end{eqnarray}
In this limit the mass splitting between $e_1$ and $e_2$ is given by $m_{e_2} - m_{e_1} = \sqrt{2} y_e v_h$ while $m_{\nu_4} - m_{e_1} = \frac{1}{\sqrt{2}} y_e v_h$. For $y_e \sim 0.8$ this gives a mass splitting of $\sim 280$~GeV between the charged leptons and a splitting of $\sim 140$~GeV between $e_1$ and $\nu_4$. Note also that for $y_e \sim 0.8$ and the $m_{e_1}$ range $100$~GeV $-$ 500~GeV shown in Fig.~\ref{fig:LHCpair} one has $0.2 \lesssim c_e \lesssim 0.53$. The cross sections can be as large as one pb for particle masses close to the LEP limits, and up to 50~fb for particle masses in the several hundred GeV range.

The decays of the exotic leptons will lead to a number of signatures at the LHC via their decays to electroweak gauge and Higgs bosons as well as DM. In the limits leading to Eq.(\ref{eqn:DM_masses}) and Eq.(\ref{eqn:emasses}) the heavy charged state $e_2$ can have the following decay chain,
\begin{eqnarray}
\label{eqn:echain1}
e_2 \rightarrow W \nu_4 \rightarrow W W e_1 \rightarrow WWW \nu_X.
\end{eqnarray}
Note that although we are neglecting mass mixing between $\nu_X$ and $\nu_4$ by assuming $y_\nu \ll 1$, it must be non-zero for the the heavy leptons to decay down to the DM.

One can also have the heavy charged state decaying to DM more directly via,
\begin{eqnarray}
\label{eqn:echain2}
e_2 \rightarrow Wh \nu_X, ~e_2 \rightarrow WZ \nu_X, ~e_2 \rightarrow W \nu_X\,,
\end{eqnarray}
while the light charged state only has one tree level decay, 
\begin{eqnarray}
e_1 \rightarrow W \nu_X.
\end{eqnarray}
The heavy neutrino state $\nu_4$ can decay via $Z$ and $h$ bosons through,
\begin{eqnarray}
\label{eqn:signatures}
\nu_4 \rightarrow Z\nu_X,~\nu_4 \rightarrow h\nu_X ,
\end{eqnarray}
as well as $W$ bosons through,
\begin{eqnarray}
\nu_4 \rightarrow W e_1 \rightarrow W W \nu_X.
\end{eqnarray}
Thanks to the large mass differences between the particles, all intermediate gauge bosons are on-shell, such that their final states can easily be reconstructed at the LHC. These decay patterns can change in more general lepton mixing scenarios, but should offer promising channels at the LHC. 

For low masses, we see from Fig.\ref{fig:LHCpair} that $e_1^+e_1^-$ has the largest production rate. Assuming leptonic decays of the $W$-bosons, this leads to a signature
\begin{align}
	p p \to e_1^+ e_1^- \to WW {E\!\!\!/}_T \to l^+l^- {E\!\!\!/}_T \,.
\end{align}
For larger masses the $e_1^+ \nu_4$ channel becomes dominant, and can give rise to a striking trilepton signature through
\begin{align}
	p p \to e_1^+ \nu_4 \to W Z {E\!\!\!/}_T \to l^+l^+l^- {E\!\!\!/}_T \,.
\end{align}
The signatures are similar to those from production of weakly charged supersymmetric particles at the LHC. While limits can be obtained in special cases from the 8~TeV run of the LHC, we expect that at least 100 fb$^{-1}$ at the 14~TeV LHC are needed to probe the exotic lepton sector at the LHC. 

For light enough $\phi$ there is also the potential to produce it resonantly at the LHC through Higgs mixing. This scalar would inherit the SM Higgs decays, but be suppressed by $s_\theta^2$. Additionally, if kinematically allowed $\phi$ can also have the following decays to heavy leptons and dark matter,
\begin{eqnarray}
\label{eqn:phsignatures}
\phi \rightarrow e_{1} e_{1}&,&~~\phi \rightarrow e_{2}e_{2} \nonumber \\
\phi \rightarrow e_{1} e_{2}&,&~~\phi \rightarrow \nu_{4}\nu_{X} \nonumber \\
\end{eqnarray}
It can of course also decay to Higgs pairs $\phi \rightarrow hh$ when kinematically allowed. As discussed in Sec.\ref{sec:constraints}, however, for $v_\phi \sim 1.7$~TeV we typically have $\phi$ in the TeV range (see Fig.\ref{fig:lam_vs_lam}) making it phenomenologically irrelevant for much of the parameter space.

\section{Conclusions/Outlook}
\label{sec:conclusions}

We have constructed a theory based on the gauging of lepton number, and found that for many choices of the parameters,
the exotic leptons required to cancel gauge anomalies contain a dark matter candidate whose thermal relic density
naturally saturates the requirements of cosmological observation. The dark matter is a Dirac (mostly singlet) neutrino and we find that masses $\gtrsim 200$~GeV give the correct thermal relic abundance via annihilation through the massive vector boson associated with the gauged lepton number. Higgs scalar mixing as well as gauge kinetic mixing which are found in this model also allow for a direct detection signal and give reasonably good prospects for detection in near future experiments. 

The theory introduces only one new scale, the vacuum expectation value of a SM singlet scalar which breaks the lepton number and is constrained by experiment to be $\gtrsim 1.7$ TeV. The global symmetry which stabilizes the dark matter is a consequence of the gauge structure and particle content of the the theory and does not need to be additionally imposed. Furthermore, as a consequence of the lepton number breaking, the dark matter is also accompanied by a set of vector like leptons charged under the SM gauge group with couplings to the SM Higgs. The same global symmetry which stabilizes the dark matter also prevents any dangerous flavor changing neutral currents or mass mixing with SM leptons. For a lepton breaking scale $\sim 1.7$~TeV phenomenologically viable dark matter and exotic vector-like leptons can be obtained.

The model contains a variety of potential LHC signals, though rates will be challenging. Some of the signatures, such as a four lepton final state with a $Z_L$ resonance in two of the leptons are fairly novel and specific, but otherwise most LHC phenomenology resembles other vector like lepton constructions along with singlet scalar phenomenology. The 14~TeV run of the LHC should be able to probe some of the parameter space in the exotic lepton sector, although an $e^+e^-$ collider with center of mass energies between 250~GeV and 500~GeV is more suitable for this task. Unless the $Z_L$ is very light, direct production is unlikely to be observable at the LHC. The indirect effect on four lepton interactions can however be probed at a linear collider, vastly extending the reach of the LEP experiments. 

The exotic charged leptons can also lead to observable modifications of the Higgs decays and in particular to $h\rightarrow \gamma\gamma$, which is also affected by Higgs mixing. We have examined these effects for a range of model parameters and lepton masses which can potentially be produced at the LHC. Potential vacuum stability issues due to the presence of charged leptons with $\mathcal{O}(1)$ couplings to the Higgs can be alleviated with the presence of the gauge and scalar sector of this model, but one can also easily embed it into a more fundamental UV completion which would presumably solve such problems.

While $U(1)_L$ is an attractive gauge symmetry, which may contribute to the answer as to how dark matter can be massive and yet remain stable, many open questions remain in the current construction.  For example, the hierarchy problem remains unaddressed, and almost certainly would require more structure and would lead to new phenomena. The current construction automatically contains new massive states as well as new interactions potentially containing $CP$-violating phases, which may be useful for explaining the baryon asymmetry of the Universe. One can also easily imagine embedding this model into a supersymmetric version or some other construction which solves the hierarchy problem or generates the lepton breaking scale naturally, but we leave these possibilities to a future study.

\section*{Acknowledgements}

The authors thank Andr\'{e} de Gouv\^{e}a, Bogdan Dobrescu, Patrick Fox, Roni Harnik, Carlos Wagner, and Felix Yu for useful conversations.
The research of T.M.P.T. is supported in part by NSF grant PHY-0970171 and by the University of California, Irvine through
a Chancellor's fellowship. R.V.M. is supported by the Fermilab Graduate Student Fellowship program. This research is also partially supported by Fermi Research Alliance, LLC under Contract No. De-AC02-07CH11359 with the United States Department of Energy. Work of P.~S. is supported in part by the U.S. Department of Energy, Division of High Energy Physics, under grant numbers DE-AC02-06CH11357 and DE-FG02-12ER41811.

\bibliographystyle{apsrev}
\bibliography{GaugedL}

\end{document}